\documentclass[iop]{emulateapj}

\usepackage{epsfig}

\newcommand{\kms}{km s$^{-1}$}

\newcommand{\ie}{\emph{i.e.}}
\newcommand{\eg}{\emph{e.g.}}
\newcommand{\vmax}{\mathrm{v_{max}}}
\newcommand{\vmaxacc}{\mathrm{v_{max}^{acc}}}

\shorttitle{DEEP2 MOCK CATALOGS}
\shortauthors{Gerke et al.}

\begin{document}
\title{Improved Mock Galaxy Catalogs for the
  DEEP2 Galaxy Redshift Survey from Subhalo Abundance and Environment Matching}

\author{Brian F. Gerke\altaffilmark{1,$\ast$}, 
Risa H. Wechsler\altaffilmark{1,2}, 
Peter S. Behroozi\altaffilmark{1,2},
Michael C. Cooper\altaffilmark{3}, 
Renbin Yan\altaffilmark{4}, 
Alison L. Coil\altaffilmark{5}}  
\altaffiltext{1}{Kavli Institute for Particle Astrophysics and Cosmology, 
SLAC National Accelerator Laboratory, M/S 29, 2575 Sand Hill Rd.,
Menlo Park, CA 94025}
\altaffiltext{2}{Department of Physics, Stanford University, Stanford,
CA, 94305}
\altaffiltext{3}{Center for Galaxy Evolution, Department of Physics and
  Astronomy, University of California--Irvine, Irvine, CA 92697}
\altaffiltext{4}{Center for Cosmology and Particle Physics, Department of Physics, New York University, 4 Washington Place, New York, NY 10003}
\altaffiltext{5}{Center for Astrophysics and Space Sciences,
  University of California, San Diego, 9500 Gilman Dr., MC 0424, La
  Jolla, CA 92093}
\altaffiltext{$\ast$}{Current address: Energy Efficiency Standards Group, Lawrence Berkeley National Laboratory, 1 Cyclotron Rd., M/S 90R4000, Berkeley  CA 94720}
\email{bgerke@slac.stanford.edu}

\begin{abstract}
  We develop empirical methods for modeling the galaxy population and
  populating cosmological N-body simulations with mock galaxies
  according to the observed properties of galaxies in survey data.  We
  use these techniques to produce a new set of mock catalogs for the
  DEEP2 Galaxy Redshift Survey based on the output of the
  high-resolution Bolshoi simulation, as well as two other simulations
  with different cosmological parameters, all of which we release for
  public use.  The mock-catalog creation technique uses subhalo
  abundance matching to assign galaxy luminosities to simulated
  dark-matter halos.  It then adds color information to the resulting mock
  galaxies in a manner that depends on the local galaxy density, in
  order to reproduce the measured color-environment relation in the
  data.  In the course of constructing the catalogs, we test various
  models for including scatter in the relation between halo mass and
  galaxy luminosity, within the abundance-matching framework.  We find
  that there is no constant-scatter model that can simultaneously
  reproduce both the luminosity function and the autocorrelation
  function of DEEP2.  This result has implications for
  galaxy-formation theory, and it restricts the range of contexts in
  which the mocks can be usefully applied.  Nevertheless, careful
  comparisons show that our new mocks accurately reproduce a wide
  range of the other properties of the DEEP2 catalog, suggesting that
  they can be used to gain a detailed understanding of various
  selection effects in DEEP2.
\end{abstract}
\keywords{galaxies: evolution --- galaxies: high-redshift ---
  large-scale structure of the universe --- galaxies:halos --- dark matter}


\section{Introduction}
\label{sec:intro}
The accurate interpretation of galaxy survey data requires a careful
consideration of a wide range of biases and incompleteness that can
arise from the selection of the surveyed galaxies.  Galaxy selection
most commonly occurs as a function of galaxy flux, color, and
(particularly for spectroscopic surveys) local density on the sky.
Since these three properites are well known to be strongly correlated
with one another \citep[\eg][]{Hogg04, Cooper06a}, a selection
algorithm that chooses galaxies based on one of them will also produce
selection biases in the distributions of the others.  Estimating the
scale of secondary selection effects like these requires a detailed
understanding of the interplay between different galaxy properties;
this becomes more important and more complex when considering surveys
that cover a wide range in redshift, since the galaxy population may
evolve substantially over the lookback time of the survey.  The use of
mock (\ie, simulated) galaxy catalogs to understand and account for
selection effects has therefore been an important part of the analysis
in nearly all modern galaxy surveys, particularly those probing
redshifts of order unity, such as DEEP2 \citep{Davis03, DEEP2}, VVDS
\citep{VVDS}, and zCOSMOS \citep{zCOSMOS}.  

In this study, we develop a set of empirically based techniques for
constructing mock galaxy catalogs for high-redshift surveys.  We also
produce a new set of catalogs for the DEEP2 survey and make them
available to the public, although our techniques should be generally
applicable to any spectroscopic galaxy survey. One of our primary
goals in constructing the mock catalogs presented here is to produce a
simulated catalog appropriate for testing and optimizing algorithms to
detect groups and clusters of galaxies in redshift space in DEEP2.  We
also wish to use these mocks to calibrate the halo-mass selection
function that corresponds to the observational selection of galaxy
groups (\ie, systems with two or more observed members in DEEP2).

At present, the detailed astrophysics of galaxy formation is not
sufficiently well understood to directly simulate the galaxy
population from first principles.  However, there is strong support,
through many different lines of evidence, for a basic picture of
galaxy formation which holds that (1) all galaxies are embedded in
larger dark matter halos that comprise most of their mass, (2) above
some minimum mass threshold, all dark matter halos (whether primary
halos or subhalos within a primary) host a single galaxy at their
centers, and (3) there is a tight relation between the optical
luminosity of a galaxy and the mass of its dark matter host.  This
paradigm suggests an obvious approach to constructing mock galaxy
catalogs by assigning a galaxy properties to each halo or subhalo in a
dark-matter-only N-body simulation, using some prescription for
mapping between galaxy properties and (sub)halo properties.
Techniques for converting periodic N-body simulation volumes into the
cone-shaped geometries typical of surveys, and for including the
effects of cosmic structure evolution at high redshift, are now well
developed \citep[\eg,][]{YWC04, Blaizot05, KW07}.  What remains is to
specify a halo-galaxy connection that reproduces the observed galaxy
population with high accuracy.

One would ideally like to have a physically well-motivated method for
assigning galaxies to dark-matter halos, so numerous authors have
constructed mocks using semi-analytic models of galaxy formation
applied to the underlying N-body simulations \citep[\eg,][]{Eke04,
  KW07, Henriques12}.  At present, however, most such models fail to
accurately reproduce the color distribution of galaxies at redshifts
near unity.  An additional difficulty arises because, until quite
recently, N-body simulations of cosmologically interesting volumes
have not had sufficient particle numbers to resolve the relatively
low-mass halos and subhalos that host the faintest galaxies in modern
surveys.  \citet{YWC04} (hereafter YWC) addressed these problems by
taking an empirically based halo-model approach, using a conditional
luminosity function to construct mock catalogs for the DEEP2 redshift
survey, populating massive halos with one central galaxy and a number
of satellites depending on the halo mass, in such a way as to match
the measured DEEP2 luminosity and autocorrelation functions.  Since
the satellites had to be randomly assigned to dark matter particles in
these halos, rather than to well-defined subhalos, this approach
likely mis-estimated the spatial profiles and velocity distributions
of galaxy groups and clusters.  Given the simulations available at the
time, however, even this approach could only account for the
relatively bright galaxies probed by DEEP2 at $z\ga 0.7$; the fainter
galaxy population sampled at low $z$ was not included.

The YWC mocks also did not include information on galaxy properties
besides luminosity, although \citet{Gerke07a} were able to add color
information to these mocks according to the measured relation between
color and local galaxy density.  Their approach to color assignment
was inspired by the ADDGALS mock-catalog creation algorithm that was
used to produce mock catalogs for the Sloan Digital Sky Survey
\citep{SDSS} in \eg, \citet{Koester07a}, and which will be described
in detail by Wechsler et al. (in preparation).  Even with
the addition of colors, these mocks were not sufficient to model the
faint, low-redshift population probed by DEEP2 in one of its three
observational fields.  In addition, the YWC mocks were based on N-body
simulations whose underlying cosmological parameters are at variance
with the best fit to current data.  For these reasons, we are
motivated to improve upon the YWC efforts and construct new mock
catalogs for the DEEP2 survey.

Since the construction of the YWC mocks, improvements in both software
and hardware have allowed for a substantial increase in the mass
resolution of N-body simulations of a given volume.  Most recently,
the Bolshoi simulation \citep{Bolshoi} simulated the growth of cosmic
structure in a cubical volume 250 comoving h$^{-1}$ Mpc on a side,
resolving halos and subhalos down to masses smaller than those of the
Magellanic Clouds \citep{Busha10b} and using a set of cosmological
parameters that is in good agreement with current constraints from a
wide variety of data.  The combination of a large volume and excellent
mass resolution permit the construction of DEEP2 mock catalogs over
the full redshift and luminosity range probed by the survey, except
for a handful of very faint low-$z$ dwarfs.  Moreover, because Bolshoi
resolves subhalos down to the required mass range for DEEP2, we will
be able to assign galaxies to halos and subhalos individually, rather
than taking a halo-model based approach, which should better reflect
the phase-space distributions of galaxies in groups and clusters.

Lacking a sufficiently accurate astrophysical model of galaxy
formation, we carry on with a purely empirical approach to
mock-catalog construction.  A popular technique for assigning galaxy
luminosities, known as subhalo abundance matching, has been
demonstrated as a feasible means of reproducing both the galaxy
luminosity function and the autocorrelation function at a variety of
redshifts, under certain cosmological assumptions \citep[e.g.][]{
Kravtsov04, Tasitsiomi04, ValeOstriker04, Conroy06a, Reddick12};
we adopt this approach here and explore its applicability in more
detail.  To add color information to the resulting mock catalogs, we
expand on the environment-dependent approach of \citet{Gerke07a},
making substantial improvements to accurately model galaxies near the
DEEP2 flux limit, as well as various luminosity and color-dependent
sources of incompleteness in the survey.  We also repeat the
mock-making procedure for two other high-resolution N-body simulations
with different cosmological parameters from Bolshoi, to facilitate
tests for cosmological dependence in the DEEP2 selection function.

We proceed as follows.  The next section introduces the DEEP2 survey
and the N-body simulations.  Section~\ref{sec:algorithm} details our
methods for assigning galaxy properties to halos and subhalos, while
Section~\ref{sec:observation} describes the techniques we use to
replicate various DEEP2 selection effects in the resulting catalogs.
Section ~\ref{sec:compare} is likely the most useful portion of the
paper for users of these mock catalogs, since it makes detailed
comparisons between the mock catalogs and the DEEP2 dataset.  In the
Appendix, we describe the contents of the public DEEP2 mock catalogs
and give a brief example of their use for computing the DEEP2
halo-mass selection function.  Throughout this paper, unless otherwise
specified, distances are quoted in comoving $h^{-1}$ Mpc and absolute
magnitude values are given as $M-5\log h$.

\section{Data and Simulations}
\label{sec:datasim}
\subsection{The DEEP2 Galaxy Redshift Survey}
The DEEP2 (Deep Extragalactic Evolutionary Probe 2) Galaxy Redshift
Survey is a large spectroscopic survey of galaxies at $z\sim 1$
comprising spectra of some $50,000$ objects and $\sim 35,000$
confirmed galaxy redshifts in four observational fields covering a
total of $\sim 3$ deg$^2$ on the sky.  Complete details of the survey,
including target selection, spectroscopy, data reduction, and redshift
assignment procedures appear in \citet{DEEP2};
substantial discussion of these issues is also available in
\citet{DGN04, AEGIS}, and \citet{Willmer06} (hereafter W06).  Here, we
summarize the properties of the survey that we will need to replicate
in our mock catalogs.  

Spectroscopic targets for DEEP2 were selected from deep imaging with
the CFH12k camera on the Canada-France-Hawaii telescope
\citep{Coil04b} in the $B, R$ and $I$ bands.  Three of the four fields
each comprise three CFH12k pointings, oriented along lines of constant
declination, making a 1 deg$^2$ contiguous field with a 2 deg $\times$
0.5 deg aspect ratio.  The fourth field is the 2 deg $\times$ 0.25 deg
Extended Groth Strip (EGS), which is oriented perpendicular to the ecliptic
and has been the site of a wide variety of observations across the
full range of the electromagnetic spectrum.

The spectroscopic selection has an $R$-band apparent magnitude limit
of 24.1, and a selection in observed color-color space was also
applied to exclude galaxies at redshifts below $z\sim 0.75$, except in
the EGS, where galaxies were observed throughout color space, but with
a preference for galaxies expected to lie at $z > 0.75$.  Spectra
were obtained with the DEIMOS spectrograph \citep{Faber03} on the Keck
II telescope, which uses custom-milled slitmasks to allow multiplexed
slit spectroscopy of $\ga 100$ objects simultaneously.  The DEEP2
spectroscopic targeting algorithm tiles the DEEP2 fields with DEIMOS
slitmasks in an overlapping pattern such that most of the suitable
galaxies have at least two chances of being selected for spectroscopy
(four chances in the EGS).  Because the total slit length available is
limited, it is not possible to target all suitable galaxies for spectroscopy;
the targeting efficiency of DEEP2 is $\sim 60\%$.  

Data reduction and initial redshift estimation was performed using a
software pipeline \citep{DEEP2} that was specially designed for use
with DEEP2 DEIMOS data.  Each spectrum was then examined by eye and
the initial redshift estimate was either confirmed or corrected, or
the spectrum was rejected as a failiure.  Roughly $70\%$ of DEEP2
spectra yield successful redshifts. We estimate, however, that $\sim
15\%$ of DEEP2 targets lie beyond the DEEP2 target redshift range
(C. Steidel 2003, private communication), at $z>1.4$, where all useful
spectral features for redshift identification have shifted out of the
optical waveband.  Thus, the redshift-success rate for DEEP2 galaxies
in the appropriate redshift range is $\sim 85\%$; when combined with
the targeting rate, this gives an overall spectroscopic sampling rate
for DEEP2 of $\sim 50\%$.

\subsection{N-Body Simulations}

We use dark matter halos from three N-body simulations.  Our principal
results use the Bolshoi simulation \citep{Bolshoi}, which modeled a
250~$h^{-1}$ Mpc comoving box with $\Omega_m = 0.27$,
$\Omega_{\Lambda} = 0.73$, $\sigma_8 = 0.82$, $n=0.95$, and $h = 0.7$.
The simulation volume contained $2048^3$ particles, each with mass
$1.15 \times 10^8 h^{-1}M_\odot$, and was run using the ART code
\citep{Kravtsov97}.  Halos and subhalos were identified using the BDM
algorithm \citep{Klypin97}; see \cite{Bolshoi} for details. This
simulation's spatial resolution is a physical scale of 1~$h^{-1}$kpc.
This improves the tracking of halos as they merge with and are
disrupted by larger objects, allowing them to be followed even as they
pass near the core of the halo.  The resulting halo catalog is nearly
complete for objects down to a circular velocity of $\vmax = 55$ \kms.
Merger trees were created using the consistent merger tree code of
\cite{Behroozi11b}.

The additional simulations used are described in Table
\ref{tab:cones}.  These simulations were also run with the ART code
and analyzed in the same way as Bolshoi; we primarily use them to test
the impact of varying cosmological parameters on the properties of our
mock catalogs.  All results shown use the Bolshoi simulation, with the
exception of Figure 12, which compares the halo occupation distribution
between the simulations.

\section{Mock-catalog creation algorithm}
\label{sec:algorithm}
A brief summary of our algorithm for constructing DEEP2 mock catalogs
from the N-body simulations is as follows.  We first construct
redshift-space lightcones with the geometry of the DEEP2 by defining
an observer position and line-of-sight direction in the $z=0$
simulation output and then stacking different simulation outputs along
the line of sight, making use of the boxes' periodic boundary
conditions.  We then assign monochromatic ($B$-band) galaxy
luminosities to the dark matter halos and subhalos, according to the
measured DEEP2 luminosity function, using the subhalo abundance matching technique
\citep[\eg][]{Conroy06a, Behroozi10}, and we account for the measured
redshift evolution in this luminosity function.  We assign each mock galaxy
a $U-B$ color drawn from the DEEP2 survey according to its local
density, by matching the color distribution in quintiles of local
density between DEEP2 and the mock catalog, and we apply an inverse
$k$-correction to these galaxies to obtain an observed $R$-band
apparent magnitude for each one.  Finally, we apply the various DEEP2
selection cuts, including density-dependent incompleteness arising
from the scheduling of galaxies for DEIMOS spectroscopic observation
and color, luminosity, and redshift-dependent incompleteness arising
from the failure to obtain reliable redshifts for some targets.  We
explain each of these steps in detail in the following sections.

\subsection{Constructing the lightcone}

To construct a realistic mock catalog, it is necessary to convert the
real-space positions of halos and subhalos in an N-body simulation
into redshift-space positions in a lightcone that conforms to the
survey geometry.  To the extent possible, we would also like the
positions at each redshift to correspond to the appropriate epoch of
the simulation---so that, for example, mock galaxies at $z=1$ are
drawn from a snapshot of the simulation at a scale factor $a=0.5$.
Since we would like our lightcones to represent the DEEP2 survey
volume as realistically as possible, we wish to avoid duplicating any
region of the simulation volume in any one lightcone.  Numerous
authors have implemented lightcone-making for mock catalogs (e.g.,
YWC, \citealt{Blaizot05, KW07}).  Here, in the two smaller simulation
boxes (120 and 160 Mpc), we closely follow the approach of YWC.  In
the Bolshoi simulation, owing to the larger volume available, we take
a somewhat more flexible approach.  We outline these
lightcone-construction algorithms briefly below.

The basic idea is to choose an observer position within the simulation
box, an observational field geometry, and a direction of observation.
We can then trace a cone through the box, making use of the periodic
boundary conditions of the simulation to ``wrap'' the lightcone and
construct a line of sight that is significantly longer than the
nominal dimensions of the box.  Once the cone is constructed, it is
straighforward to convert the positions of halos and subhalos in the
simulation box into RA, Dec and line-of-sight distance coordinates in
the lightcone.  We then convert the line-of-sight distance into a
redshift for each object using the distance--redshift relation for the
background cosmology of the mock and adding a peculiar-velocity
contribution derived from the velocity component of each object along
the chosen line of sight.  In each simulation, we include halos and
subhalos down to masses of a few times $10^9 M_\odot$, which includes
the mass range of the faint ($M_B > -15$) dwarf galaxies that DEEP2
probes at low redshift in the EGS.  In some simulations, including
Bolshoi, the halo population is incomplete at these masses, but
low-mass dwarfs only appear at very low redshift in the EGS field of
DEEP2, and this part of DEEP2 probes an exceedingly small volume. Such
incompleteness will thus only have a very minor impact on projection
effects, and will not impact any cosmological comparisons.

It is important to take some care in the choice of line-of-sight
direction, angling it to ensure that the lightcone does not overlap
itself on subsequent passes through the simulation box and repeatedly
sample the same volume (see, \eg, the bottom panel in Figure 2 of
\citealt{Blaizot05}).  This also imposes a practical limit on the
redshift range that a given lightcone can probe, which depends on the
geometry of the observational field: a wide lightcone will
quickly reach a comoving transverse size similar to the box size for
example, making overlaps inevitable.  Since DEEP2 is nearly a
pencil-beam survey, however (the fields are $\sim 2$ degrees across in
the longest dimension), this is not a limitation for our
purposes in practice.

YWC used a simulation whose output timestep spacing
corresponded to the light-travel time over roughly half of the
simulation box dimensions.  Their approach to constructing lightcones
was to start with an observer position on one face of the simulation
cube and transition between timesteps each time the line of sight had
traversed half of the box.  It is then possible to make a second
lightcone by translating the observer position halfway across the box
in the direction perpendicular to the face that was chosen initially.
This procedure can be repeated for each of the six box faces, allowing
a maximum of twelve lightcones to be constructed.  If more lightcones or
a finer time-spacing are desired, it is possible to  generalize this
approach to start with a randomly selected observer location and
sightline, subject to the requirement that the lightcone not overlap
itself \citep[\eg][]{KW07}.   

How many lightcones should we produce for each simulation?  Since we
would like to use our mock catalogs to estimate the sample variance in
DEEP2, we would like our lightcones to probe independent (\ie,
nonoverlapping) volumes as much as is possible.  In the ideal case,
all of our mock lightcones would contain completely independent
volumes, and so we should stop constructing lightcones once their
total volume equals the volume of the simulation box\footnote{Neither
  of our algorithms absolutely guarantees that two different
  lightcones will not overlap, of course, but since DEEP2 is nearly a
  pencil-beam survey, any overlapping volume between mocks should be
  small.}.  For many purposes, however, it is sufficient if the mocks
are independent only at fixed redshift, or, more precisely, in
redshift bins of the width that we are typically interested in
considering in our analyses.  That is, if we want to estimate the
sample variance in bins of width $\Delta z = 0.1$, then it is
acceptable if mock $A$ contains the same volume at $z=1$ that mock $B$
contains at $z=0.5$.  For this reason, we will tend to ``overfill''
our simulation boxes with mock catalogs by a factor of a few, to allow
for better statistical power in estimating the sample variance in
redshift bins.

A single $1\; \mathrm{deg}^2$ DEEP2 observational field contains a total
volume of $\sim 2\times 10^6 h^{-3} \mathrm{Mpc}^3$ over the primary
DEEP2 redshift range of $0.75<z<1.4$, and a redshift bin of width
$\Delta z = 0.1$ has a volume $\sim 3 \times10^5
h^{-3}\mathrm{Mpc}^3$.  Thus, we can fit one complete DEEP2 field into
the L120 simulation, and two into the L160 box, and on the order of
ten independent $\Delta z \sim 0.1$ redshift bins into each.  We
therefore use exactly the algorithm that was used in YWC to construct
lightcones for these two boxes, producing twelve lightcones for each
box.  Given the total volume of the boxes, we expect that redshift
bins of width $0.125$ will contain independent volumes across all
twelve cones in the L160 box, and bins of width $0.05$ will be
independent across the L120 lightcones. These numbers are summarized
in Table~\ref{tab:cones}.  In each case, wider bins will allow
proportionally fewer independent cones.  Both of these simulations
have sufficient time resolution in their outputs that we can follow
the YWC technique of transitioning between timesteps whenever the
sightline has traversed half the box---except in the case of the L120
box at $z>1$.  In that regime, there are only a few simulation outputs
covering the entire range from $1.0<z<1.4$.  For this reason, we
advise against using the L120 box for applications that require an
accurate representation of the redshift evolution of cosmic structure.

 The larger volume of the $250 h^{-1} \mathrm{Mpc}$ Bolshoi simulation
 allows us to construct significantly more independent DEEP2 samples,
 with the volume of eight fully independent DEEP2 fields (over the
 primary $0.75<z<1.4$ redshift range) available in Bolshoi, and scores
 of independent $\Delta z \sim 0.1$ redshift bins available.  We
 therefore use the more flexible approach to lightcone-building
 described above, with randomly selected observer positions and lines
 of sight, to construct forty DEEP2 lightcones from the Bolshoi
 simulation.   These forty lightcones should be almost completely
 independent in redshift bins as wide as $0.15$.  The time resolution
 of Bolshoi is sufficiently good over the entire redshift range
 sampled to allow multiple timesteps to be used in each crossing of
 the box, so the growth of large-scale structure should be captured
 with sufficient accuracy for any practical purposes.  

\begin{deluxetable*}{lcccccccc}
\tablecolumns{9}
\tablewidth{0pc}
\tablecaption{Parameters of the three N-body simulations used to
  produce mock catalogs in this work.}
\tablehead{
\colhead{Simulation}   & \colhead{Box Dimension} & \colhead{$\Omega_M$}
& \colhead{$\sigma_8$} & \colhead{$h$} &\colhead{$\#$ of} &
  \colhead{$\#$ independent} & \colhead{Max. independent} \\
\colhead{} &  \colhead{($h^{-1}$ Mpc)} & \colhead{} & \colhead{} &
  \colhead{}& \colhead{lightcones} &\colhead{DEEP2 fields} 
&\colhead{$z$ bin width\tablenotemark{a}}
}
\startdata
L120 & 120 & 0.3 & 0.7 & 0.73 & 12 & 1 & 0.05 \\
L160 & 160 & 0.24 & 0.9 & 0.7 & 12 & 2 & 0.15 \\
Bolshoi & 250 & 0.27 & 0.82 & 0.7 & 40 & 8 & 0.15 \\
\enddata
\tablenotetext{a}{So that, for example the same redshift bin in the
  L160 lightcones samples a different region of space in all twelve
  cones.}
\label{tab:cones}
\end{deluxetable*}

\subsection{Luminosity assignment by subhalo abundance matching}
\label{sec:abundancematch}

As discussed in the introduction, one of our primary goals is to
produce a catalog that can be used for testing and calibrating
group-finding algorithms. Therefore, we want to ensure that our mock
catalogs accurately reproduce both the \emph{number} of observed
galaxies in massive halos and their \emph{distribution} in phase
space, since redshift-space cluster-finding algorithms will be
strongly sensitive to both of these.

\subsubsection{Subhalo abundance matching: rationale and basic
  approach}

In the language of the halo model, our first requirement translates to
accurately reproducing the halo occupation probability, $P(N|M)$,
which is the probability that $N$ galaxies brighter than some
luminosity threshold dwell in a halo of mass $M$.  This distribution
is typically represented by its first moment, $\bar{N}(M)$, typically
called the halo occupation distribution (HOD), plus an assumed form
for the scatter in $N$ (see, \eg, \citealt{Zheng05}).  For a given
background cosmology, a particular choice of HOD uniquely specifies
the two-point autocorrelation function of galaxies \citep{PS00}.  If
we assume our simulation has the correct background cosmology, we can
choose a prescription for populating it with mock galaxies and test
the accuracy of the HOD by comparing its correlation function to the
one measured in the DEEP2 data.  YWC took this basic approach: they
assigned galaxies to massive dark matter halos as a function of mass
according to a conditional luminosity function (CLF; see
e.g. \citealt{Yang03}), $\phi(L|M)$.  It was possible to tune the CLF
to reproduce the observed DEEP2 galaxy clustering and, once this was
achieved, to have some confidence that the resulting HOD in the mocks
was an accurate representation of the DEEP2 HOD\footnote{However, the
  comparison between the two-point functions of the mocks and the data
  was performed using early DEEP2 measurements with large
  uncertainties.  More recent measurements of the DEEP2 clustering are
  not consistent with the YWC mocks, sharpening the need for a new set
  of catalogs.  In addition, the background cosmology in the YWC mocks
  is no longer consistent with present data.}.

However, the N-body simulations YWC used did not have sufficiently
high resolution to include the low-mass subhalos that would be
expected to host the faintest DEEP2 galaxies in groups.  Thus, instead
of placing all galaxies in halos or subhalos, those authors assigned a
single galaxy to the center of each halo and satellite galaxies to the
positions and velocities of dark-matter particles selected at random
from the halo.  Since the distribution of galaxies within a cluster is
unlikely to perfectly trace the dark matter distribution, this
approach will produce inaccuracies in the phase-space distribution of
galaxies on small spatial scales.  Put another way, the YWC mocks do
not include the effects of luminosity-dependent galaxy bias (except to
the extent that the brightest galaxy in each halo was placed at the
center).  This could have important implications for cluster finding
if bright galaxies in clusters have a significantly different spatial
or velocity distribution than faint ones. (Indeed, \citet{Coil06a}
showed that radial distribution of galaxies in halos in the YWC mocks
was not consistent with the DEEP2 data.)

For this reason, we have been careful in this work to choose
high-resolution N-body simulations in which the lowest-mass halos and
subhalos have a number density that matches or exceeds that of the
faintest galaxies observed in DEEP2.  This will permit us to assign
all galaxies directly to the centers of halos and subhalos and thus,
presumably, to more accurately reproduce the phase-space
distribution of DEEP2 galaxies.  This also allows us to bypass the
step of selecting a CLF or HOD and tuning it to
match the DEEP2 correlation function.  Instead, we can construct a
direct relation between a galaxy's luminosity and the properties of
its host halo or subhalo, choosing this relation in such a way as to
reproduce the DEEP2 clustering results.  

The simplest way to construct such a relation is to match the
luminosity function of galaxies directly to the mass function of
subhalos at fixed number density, which is the core of the abmdance
matching technique.  More precisely, we integrate the luminosity function
to compute the number density of galaxies brighter than some
luminosity, $n(>L) = \int_L^\infty \phi(L) dL$ and compare it to the
number density of halos and subhalos more massive than some mass,
$n(>M)$ (under a particular choice of halo-mass
definition)\footnote{It is worth noting here that we always compute
  $n(M)$ over the entire simulation box, rather than for the
  lightcones individually, to minimize the effects of sample variance
  on our abundance-matching procedure.}.  Setting $n(>L) = n(>M)$ then
yields an implicit relation between luminosity and mass, $L(M)$ that
must obtain if a galaxy's luminosity depends only on the mass of its
host halo.  In reality, although there is strong evidence for a tight
relation between galaxy luminosity and halo mass, there are numerous
other factors that can affect galaxy luminosity at fixed mass, so some
scatter in this relation is to be expected.  In any event, a
particular implicit mean $L(M)$ relation and scatter about that
relation will correspond to a particular HOD, and hence a particular
galaxy autocorrelation function, since the subhalos of massive halos
will be populated by more or fewer bright galaxies depending on the
details of the chosen relation.  If we assume that the basic picture
of a tight $L(M)$ relation with some scatter is correct, then we can
extract this relation from the data by populating subhalos in a
simulation according to such a relation and varying the details (\eg,
the size of the scatter or the subhalo mass definition) until we
obtain an autocorrelation function in the simulation that matches the
measured one.

\citet{Conroy06a} showed that the basic features of the measured
autocorrelation function could be reproduced using this technique,
both at low redshifts, from SDSS, and at high redshifts, from DEEP2,
given two minor alterations to the basic algorithm described above.
First, they used the maximum circular velocity values, $\vmax$, of the
halos, rather than their masses, to compute their number density
function ($\vmax$ is a more robust tracer of halo mass than the total
mass of the particles in the halo, because it traces the central part
of the halo which is less sensitive to tidal stripping
\citep{Kravtsov04, Muldrew11, Knebe11}).  In addition, when a halo was
in fact a subhalo of a more massive object, they assigned it the
$\vmax$ value it had \emph{at the time it was accreted into the larger
  halo}, $\vmaxacc$, for the purposes of computing halo number
density.  That is, they used abundance matching to derive an implicit $L(\vmaxacc)$
relation.  The choice of the accretion-time $\vmax$ amounts roughly to
assuming that satellite galaxies falling into larger groups or
clusters of galaxies are stripped of their outer dark matter halos but
not stripped of stars, so that after they are accreted their
luminosities are higher than the mean $L(\vmax)$ relation for
distinct halos.  \citet{Conroy06a} found that ignoring this effect led
to a too-low correleation function on small scales.

\subsubsection{Testing the impact of scatter}
\label{sec:scatter}
\begin{figure*}
\centering
\includegraphics[width = 5in, angle=90]{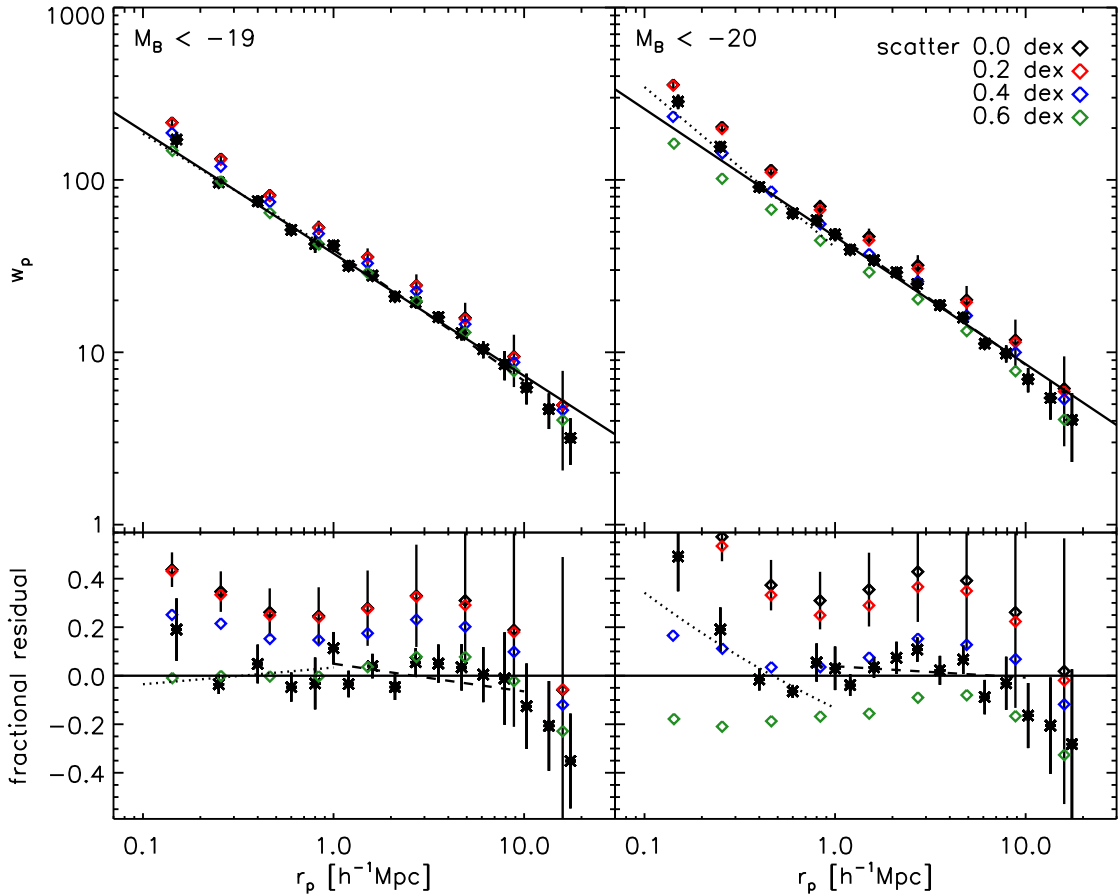}
\caption{Comparison of the projected autocorrelation functions,
    $w_p(r_p)$ in DEEP2 and in our abundance matching populations of the Bolshoi
    simulation with different amounts of scatter, for two different
    absolute magnitude thresholds.  Black points show the DEEP2
    clustering results \citep{Coil06b}, and the solid, dashed, and
    dotted lines show the best-fitting power laws to those data for
    all scales and scales greater than and less than 1 Mpc,
    respectively.  The colored points denote the clustering measured
    in the Bolshoi simulation at $z\sim 0.9$ after abundance-matching
    to the DEEP2 $B$-band luminosity function, assuming various
    amounts of scatter in the $L(\vmaxacc)$ relation, as shown in the
    legend.  Error bars on the colored points have been suppressed for
    all but one of the scatter values to reduce clutter.  There is
    little significant effect on clustering for scatter below 0.2 dex,
    and all such models are more strongly clustered than the DEEP2
    data. Higher values of the scatter produce more significant
    impacts on the clustering, but no value of the scatter is
    consistent with the DEEP2 data at all luminosity thresholds.}
\label{fig:wprp_comparison}
\end{figure*}

Introducing scatter into the $L(\vmaxacc)$ relation generally has the
effect of reducing the spatial correlation of the mock galaxies.  This
is because the mass function is steeply falling, so that there is
always a much larger population of lower-mass halos than higher-mass
ones for a given reference mass, so scatter in luminosity at fixed
mass will include a large number of low-mass halos in the population
brighter than a given luminosity, compared to the scatter-free case.
Since halo bias (the ratio of the halo correlation function to the
dark-matter one) is an increasing function of mass, including low-mass
halos will suppress the clustering.  Reddick et al (in preparation)
found that the subhalo abundance matching approach yields the best match to the SDSS
autocorrelation function and conditional luminosity function if the
$L({\mathrm v_{peak}})$ relation used has a scatter in luminosity at
fixed halo circular velocity (using the maximum value of this circular
velocity over each halo's history, ${\mathrm v_{peak}}$), with a
scatter of $\sim 0.18$ dex yielding the best results.

To investigate the appropriate scatter to use for reproducing the DEEP2
autocorrelation function, we choose the output of the Bolshoi
simulation at $a=0.528$ (corresponding to $z=0.9$, near the peak of
the DEEP2 redshift distribution) and use abundance matching to populate its halos
and subhalos with galaxies drawn from the $z=0.9$ DEEP2 luminosity
function measured by W06.  We repeat this exercise with various
different amounts of log-normal scatter in the implicit $L(\vmaxacc)$
relation, from $0$ to $0.6$ dex; in each case the scatter is constant
as a function of $\vmaxacc$.  We then compute the projected
correlation function, $w_p(r_p)$, as viewed along one axis of the box,
and we compare with the measured values in DEEP2 \citep{Coil06b}, for
galaxies brighter than a two different thresholds in luminosity.

The scatter model is implemented by an iterative non-parametric
forward convolution method, as in \cite{Behroozi10}.  A fiducial guess
for the deconvolved median $L(\vmaxacc)$ relation is obtained by
abundance matching the luminosity function to halos with zero scatter.
This relation is then convolved with the desired scatter model to
generate fiducial halo luminosities.  Abundance matching is repeated;
instead of matching the brightest galaxy to the halo with the greatest
$\vmaxacc$, however, the brightest galaxy is matched to the halo with the
highest fiducial luminosity. This in turn generates a new fiducial
median $L(\vmaxacc)$ relation, which can again be convolved with the
scatter model to generate new fiducial luminosities.  After several
iterations, this deconvolution approach converges to a stable relation
for the underlying median $L(\vmaxacc)$ relation if it exists, along with
the distribution of halo vaccmax with luminosity (i.e., $P(\vmaxacc|L)$).

Because no assumptions about the scatter model are made in this
process, we may explore a variety of models.  However, not all scatter
models are ``allowed'' in the sense that the steep fall-off of the
luminosity function imposes a stringent limit on the scatter for
luminous galaxies.  Even if the underlying relation for $L(\vmaxacc)$
imposed a sharp maximum limit on the luminosity, the addition of
log-normal scatter would broaden the fall-off in the luminosity
function; for DEEP2, this broadening becomes inconsistent with the
observed fall-off for constant scatter models above 0.3 dex.  In these
circumstances, our implementation recovers a solution for
$L(\vmaxacc)$ which, when convolved with the scatter model, cannot
match the luminosity function at the luminous end, and so only
approximately matches the luminosity function at lower luminosities as
well.

Figure~\ref{fig:wprp_comparison} shows the comparison for absolute
magnitudes $M_B \le -19$ and $M_B \le -20$.  Open data points denote
the projected correlation functions we obtain from the
abundance-matched Bolshoi simulation with different assumed values for
the scatter. Error bars are computed using jackknife sampling in the
box; hence they include the impact of cosmic variance on scales
smaller than the box (in the Figure they have been suppressed for all
but the zero-scatter model, to reduce clutter).  Black solid data
points show the $w_p(r_p)$ measurements of \citet{Coil06b}.  The solid
line shows the best-fitting power representation of the clustering
obtained over all scales in that work, and the dashed and dotted lines
show the best fits on scales larger and smaller than 1 Mpc,
respectively.  The lower panels show the fractional difference between
the best fit to the data (solid line) and the results of the
abundance-matching exercise with different amounts of scatter.

Substantial scatter appears to be needed to reproduce the DEEP2
clustering results to good accuracy: for scatter values below 0.4 dex,
the abundance-matched Bolshoi catalog is significantly more clustered
than DEEP2 for all threshold luminosities.  This suggests that the
results of \citet{Conroy06a}, who matched the DEEP2 clustering with a
zero-scatter abundance matching model, were an accident of their having used a
simulation whose background cosmology had a high value of
$\sigma_8$.  For higher values of the scatter, we find that it is
possible to reproduce the DEEP2 two-point function for a given
threshold luminosity, with a scatter of 0.4 dex beign sufficient for
$M_B<-20$ and 0.6 dex working well for $M_B<-19$, but there is no
fixed-scatter model that reproduces the \citet{Coil06b} results for
all threshold luminosities.  We note in passing that this outcome is
in qualitative agreement with the study by \cite{Wetzel10}, who found
that a scatter of $\sim 0.6$ dex was required to approximately match
the DEEP2 clustering results with an abundance matching approach.  Nevertheless, the
lack of a single scatter value that consistently matches the DEEP2
clustering at all luminosity thresholds suggests that this 
fixed-scatter abundance matching approach may be too simplistic to account for the
true galaxy population at high redshift, at least for samples selected
in the B-band.  

\begin{figure}
\centering
\includegraphics[width = 3.0in]{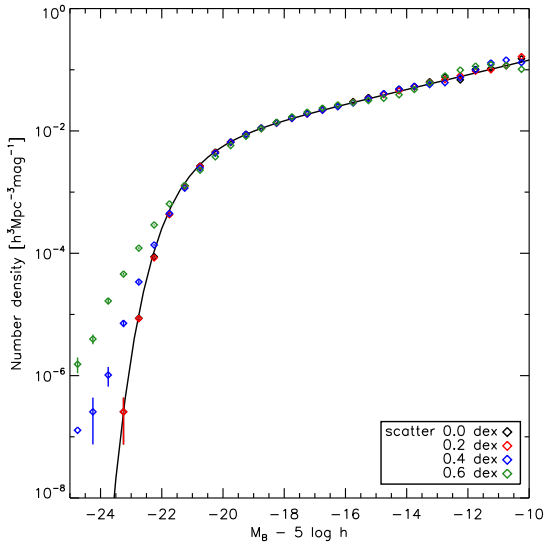}
\caption{A comparison of the input DEEP2 luminosity function (solid
  line) to the actual luminosity functions achieved in the various
  scatter scenarios for abundance matching shown in
  Figure~\ref{fig:wprp_comparison}.  The error bars shown for the mock
  scenarios reflect Poisson uncertainty.  For scatter of $\la 0.2$ dex
  in luminosity at fixed $\vmaxacc$, the input luminosity function is
  reproduced accurately, but for larger values of the scatter, the
  output catalog contains too many bright $L>^\ast$ galaxies,
  indicating that such large values of the scatter in a fixed-scatter
  abudance matching approach are inconsistent with the DEEP2
  luminosity function, given the Bolshoi cosmology. }
\label{fig:lf_comparison}
\end{figure}

In fact, the inconsistency is worse than it appears in
Figure~\ref{fig:wprp_comparison}.  As mentioned above, our technique
for adding scatter to the abundance matching approach does not
necessarily guarantee that the output mock galaxy catalog will match
the desired luminosity function. Figure~\ref{fig:lf_comparison} shows
the luminosity function we obtain from the model Bolshoi catalog for
the same scatter models discussed above (colored points), compared to
the W06 luminosity function that we used as input to the abundance
matching algorithm.  The models with 0.2 dex scatter and below match
the input luminosity function well, but models with higher scatter
deviate significantly from the W06 function at bright luminosities.
The reason for the discrepancy is straightforward to understand.  The
output luminosity function from our abundance matching algorithm is
effectively a convolution between the log-normal scatter and an
implicit no-scatter luminosity function, which must necessarily be
steeper at the bright end than the input DEEP2 luminosity function,
which is itself already quite steep.  At some value for the scatter,
the required no-scatter luminosity function will have an infinite
slope at the bright end, and for larger values of the scatter it will
be impossible to reproduce the DEEP2 luminosity function.

Figure~\ref{fig:lf_comparison} therefore suggests that the maximum
fixed log-normal scatter for abundance matching that is consistent
with the W06 DEEP2 luminosity function is $\sim 0.2$ dex in luminosity
at fixed $\vmaxacc$.  Given that no such model is consistent with the
DEEP2 correlation function, we conclude that the simple fixed-scatter
abundance matching technique for connecting galaxies with dark-matter
halos is inconsistent with DEEP2 data in detail. It is worth noting
that the DEEP2 luminosity function is measured in the rest-frame $B$
band, which is sensitive to transient effects like starbursts and AGN,
whereas the SDSS luminosity function that has been reproduced by
fixed-scatter abundance matching is measured in the $r$ band, which
should be less sensitive to such effects.  In any case, the fact that
the DEEP2 luminosity-function discrepancy is limited to bright
magnitudes, coupled with the varying amounts of scatter required to
match the clustering for different luminosity thresholds, suggest that
a variable-scatter abundance matching model whose scatter decreases as
a function of $\vmaxacc$ might yield consistency will all of the DEEP2
measurements.  One such model has been recently described by
\cite{TrujiloGomez11}, which forces agreement with the input
luminosity function at the expense of non-constant scatter.  This
model is not a unique solution, and the possible form and parameters
of models with non constant scatter are unclear, so a detailed
exploration of these issues is outside the scope of this work.

\begin{figure}
\centering
\includegraphics[width = 3.5in]{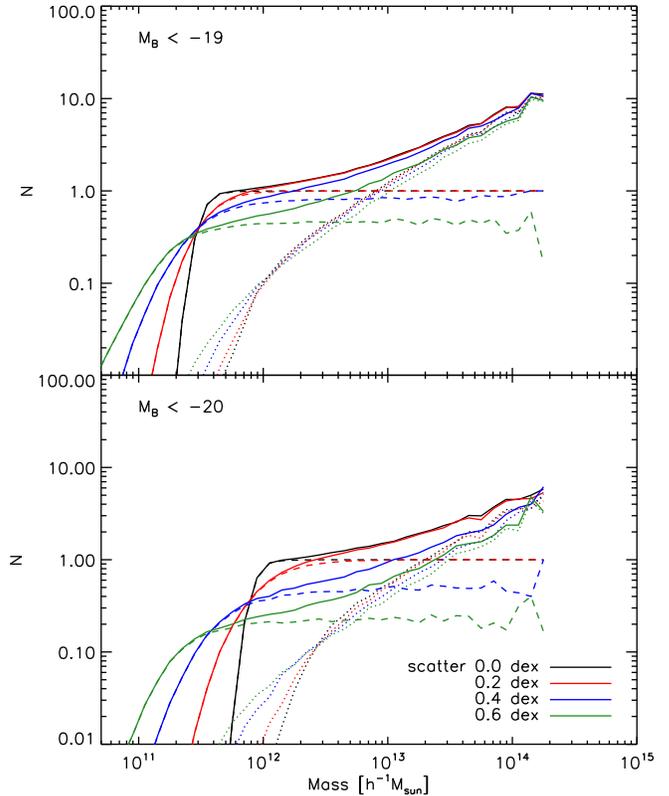}
\caption{The halo occupation distributions (HODs) that result from
  applying each of the scatter values for abundance matching in
  Figure~\ref{fig:wprp_comparison} to the Bolshoi simulation.  Solid
  lines show the mean number of galaxies per halo $\bar{N}(M)$, and
  dashed and dotted lines show the mean numbers of central and
  satellite galaxies, respectively.  For scatter values larger than
  $0.2$ dex in luminosity at fixed $\vmaxacc$, the mean number of
  central galaxies remains below unity at all masses, which is
  radically inconsistent with current understanding of galaxy
  formation.  This is a further indication that such large values of
  the scatter in abundance matching are  problematic.}
\label{fig:hod_scatter}
\end{figure}

In order to proceed, then, we must choose between an accurate
luminosity function and an accurate two-point function. To guide our
choice, we consider that our primary goal in constructing mock
catalogs is to reproduce the observational selection effects in the
DEEP2 sample as accurately as possible.  From that perspective, an
inaccurate luminosity function is likely to be more damaging than
an inaccurate two-point function, so we choose to prioritize the
former over the latter.  Only models with scatter below $\sim 0.2$ dex
are then allowed.  As shown in Figure~\ref{fig:wprp_comparison},
scatter values in this range have little to no significant impact on
the clustering, so in the interest of simplicity in our modeling, we
choose to implement a zero-scatter abundance matching model to construct DEEP2 mock
catalogs.

Before we move on, however, it is worth exploring the impact that
including scatter would have on the mocks we produce.
Figure~\ref{fig:hod_scatter} shows the HODs that result from the
various fixed-scatter abundance matching models we considered in the previous two
figures, for two different luminosity thresholds.  Solid lines show
the overall HOD $\bar{N}(M)$, dashed lines show the HOD for central
galaxies, $\bar{N}_c(M)$, and dotted lines show the HOD for
satellites, $\bar{N}_s(M)$.  The first thing to note is that, for
scatter values above 0.2 dex, $\bar{N}_c$ never reaches unity,
implying that even some very massive halos do not host bright central
galaxies in these models.  This occurs because, as mentioned above,
the no-scatter luminosity function has become infinitely steep at the
bright end, which is equivalent to the mean $L(\vmaxacc)$ relation's
becoming flat at high $\vmaxacc$.  In that case, the large scatter
will cause some central galaxies to scatter below the luminosity
threshold of interest at all masses.   This situation is strongly at
odds with theories of galaxy formation, however and is another
indication of the problematic nature of these high-scatter models.

For the models with scatter of 0.2 dex and below, the HODs are more
well behaved, and we note that the impact of the scatter occurs mainly
at low masses, softening the cutoff in $\bar{N}$ but having little
impact on the form of the HOD at large masses. We expect that a more
accurate, variable-scatter abundance matching model would increase the
scatter in luminosity at low subhalo masses, while keeping the scatter
small at high masses.  This would be expected to further soften the
low-mass cutoff, while having similarly little impact on the power-law
part of the HOD at high masses.  Thus, we anticipate that the galaxy
mass-selection function we infer from our final mock catalogs in this
work will be somewhat steeper in mass than occurs in the real
universe.  However, the mocks should prove reasonably accurate above
that mass threshold, and in particular they should give an accurate
representation of the galaxy occupation of massive groups and
clusters, at least for the bright galaxies observed by DEEP2, making
them appropriate for use in optimizing cluster-detection algorithms.
We explore their use for that purpose in \citet{Gerke12}.  We also
note in concluding this section that, should a more accurate,
variable-scatter modeling approach be developed in the future, it
could straightforwardly be combined with the techniques we develop
below to assign galaxy colors and produce realistic DEEP2 mock
catalogs that account for all observational selection effects.

\subsubsection{Populating the lightcones with an evolving DEEP2
  luminosity function}

To produce the galaxy luminosities in our mock lightcones, then, we
will abundance match the $\vmaxacc$ function of halos and subhalos to
the DEEP2 luminosity function, with zero scatter in the resulting
$L(\vmaxacc)$ relation.  This is complicated somewhat by the broad
redshift range covered by DEEP2, over which the luminosity function
undergoes significant evolution (\eg, W06, \citealt{Faber07}).
Fortunately, this evolution is consistent with simple luminosity
evolution of the global galaxy population, with no overall evolution
in number density or luminosity-function shape. Therefore, we can
reproduce the DEEP2 luminosity funciton by performing abundance
matching with a single luminosity function, measured at a particular
redshift $z_0$ in DEEP2, and then applying a redshift-dependent
correction to the luminosities of the mock galaxies.

\citet{Faber07} found that the evolution of the overall DEEP2
luminosity function was reasonably well described by a pure linear
brightening of $M_B^\ast$ by 1.2 magnitudes per unit increase in
redshift.  However, redshift is not particularly well motivated
physically as a parameter against which to measure galaxy evolution.
We have thus reconsidered the measured $M_B^\ast$ data points from
\citet{Faber07} as a function of the cosmic scale factor $a=1/(1+z)$.
A linear fit to the scale factor yields a better description of the
data than a linear fit to redshift; therefore, in constructing the
DEEP2 mocks we will implement an evolving $M_B^\ast$ that dims by 2.45
magnitudes per unit increase in $a$.

More
exactly, we populate the mock DEEP2 lightcones using an evolving
luminosity function that takes the \citet{Schechter76} form,
\begin{eqnarray}
\phi(M) &=& 0.4 \ln(10) \phi^\ast 10^{0.4(M^\ast(a)-M)(1-\alpha)}\\
&&\times \exp(-10^{0.4M^\ast(a)-M}), \nonumber
\label{eqn:LF}
\end{eqnarray}
with
\begin{equation}
M^\ast(a) = M^\ast(a_0) + Q_a(a-a_0)
\label{eqn:qevol}
\end{equation}
and $a_0 = (1+z_0)^{-1}$.  
The parameters we use are listed in
Table~\ref{tab:LFparams}.  They are based on the DEEP2 values measured
at $z\sim 0.9$ in W06, with a few modifications.
First, the value for $M_B^\ast$ is shifted brighter by $0.13$
magnitudes to be consistent with the best fit to the evolution of
$M_B^\ast(a)$.  Second, because the measured value of $\phi^\ast$ at
$z=0.9$ is the highest measured value in all of DEEP2, we instead take
the mean of all values measured at $z>0.7$ in DEEP2, which lowers
$\phi^\ast$ from the $z=0.9$ value by about $12\%$.   

Finally, we also correct this value of $\phi^\ast$ to account for the
different cosmological background models in each of our three N-body
simulations.  W06 assumed a $\Lambda$CDM cosmology with
$\Omega_M = 0.3$ in their measurements, using a volume element
computed in that cosmology to convert the redshift-space galaxy counts
to a number density.   Since they would have inferred a different
number density had they assumed the cosmology in each of our
simulation boxes, we must correct the measured value of $\phi^\ast$ by
a factor $V_{0.3}/V_{\Omega_M}$, where the volume is given in
terms of the comoving line-of-sight distance $r(z)$ as 
\begin{equation}
V_{\Omega_M} = 4\pi\int_{0.8}^ {1.0} r^2(z) \frac{dr}{dz} dz, 
\label{eqn:volume}
\end{equation}
$V_{0.3}$ is the value for $\Omega_M=0.3$, and the integral is taken
over the redshift range considered for the $z\sim 0.9$ measurement of
the DEEP2 luminosity function in W06.  The comoving
distance is given by the usual $\Lambda$CDM equation 
\begin{equation}  
\frac{dr}{dz} = \frac{c}{H_0\sqrt{\Omega_\Lambda + \Omega_M(1+z)^3}}.
\end{equation}

\begin{deluxetable}{lr}
\tablecolumns{2}
\tablewidth{0pc}
\tablecaption{Parameters of the evolving luminosity function used to
  produce the mock lightcones.}
\tablehead{\colhead{Parameter} & \colhead{Value}}
\startdata
$M_B^\ast - 5 \log h$\tablenotemark{a} & $-20.8$ \\
$\phi^\ast$\tablenotemark{b} &  $7.90 \times 10^{-3}$ \\
$\alpha$ & $1.30$ \\
$Q_a$ & $2.45$  \\
$z_0$ & $0.90$ \\
\enddata
\tablenotetext{a}{AB magnitudes.}
\tablenotetext{b}{In comoving $h^{3}$ Mpc$^{-3}$, for a $\Lambda$CDM
  cosmology with $\Omega_M = 0.27$.  This value is corrected to
  correspond to account for the different volume-redshift relation in
  each mock cosmology. }
\label{tab:LFparams}
\end{deluxetable}

\begin{figure}
\includegraphics[width = 2.75in, angle=90]{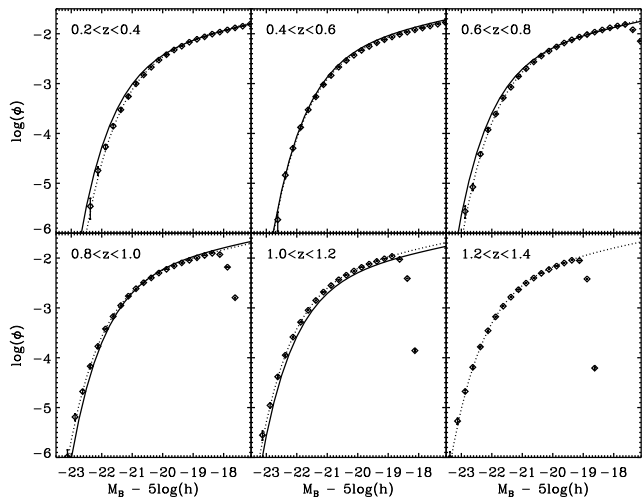}
\caption{Comparison of the DEEP2 luminosity function to the luminosity
  function in the mock catalogs.  Solid lines are the DEEP2 luminosity
  function as measured in W06.  Dotted lines are the evolving
  luminosity function as parameterized in Table~\ref{tab:LFparams},
  which was used as an input to the mock-catalog creation algorithm.
  Data points are the measured number density of galaxies in the mock,
  in bins of luminosity, with units of $h^3 \mathrm{Mpc}^{-3}
  \mathrm{mag}^{-1}$ and Poisson error bars.}
\label{fig:LF}
\end{figure}

The procedure we have outlined here ensures, by construction, that the
luminosity function of our DEEP2 mock catalogs will match the input
luminosity function to within Poisson noise.  Figure ~\ref{fig:LF}
shows the comparison in bins of redshift.  Data points are the
measured luminosity function for all 40 of the Bolshoi lightcones,
solid lines are the curves corresponding to the measured parameters
from DEEP2 in W06, and dotted lines are curves corresponding to
Equation~\ref{eqn:LF} for the parameters given in
Table~\ref{tab:LFparams}.  The dropoff in the low-luminosity
datapoints at high redshift occurs because of an evolving luminosity
cut we apply to the catalog, which excludes galaxies that are
obviously below the DEEP2 apparent magnitude limit, to keep
the catalog sizes manageable.  Apart from this, the agreement between
the luminosity function in the mocks and the input model is perfect,
as expected.

\subsection{Color assignment by environment matching}
\label{sec:envmatch}

Accurately reproducing the luminosity function, though necessary, is
not sufficient to fully model the DEEP2 galaxy selection function.  As
discussed at length in, \eg, W06 and \citet{Gerke07a}, DEEP2's
$R=24.1$ apparent magnitude limit translates into a $B$-band limit at
$z\sim 0.75$ and a $U$-band limit at $z>1$.  This means that, at a
given luminosity, red galaxies will drop out of the DEEP2 sample at a
lower redshift than blue galaxies.  There is a well-known strong
correlation between galaxy color and galaxy environment, \ie, local
galaxy density \citep[\eg][]{Hogg04}, and this relation persists to $z
> 1$ in DEEP2 \citep{Cooper06a}, although it evolves strongly with
redshift \citep{Cooper07, Gerke07a}.  If we are to use our mock
catalogs for applications that probe the galaxy density field, such as
testing and calibrating cluster-finding algorithms, it is important to
include the effects of color-dependent selection on the spatial
sampling of galaxies.  This will require that we accurately reproduce
the color-environment relation.

\citet{KW07} used a semi-analytic model for galaxy formation (from
\citealt{Croton06}, as updated by \citealt{DB07}), in an effort to
reproduce such effects in their mock catalogs.  Unfortunately, this
model does not accurately reproduce the color-magnitude-environment
relation that is present in DEEP2 at $z\sim 1$; in particular, it
produces too few faint, red galaxies in dense regions.  This is a
common feature of most semi-analytic models at present, and it likely
to be a particular problem for using these mocks to test group-finding
algorithms, since the number of observed galaxies in real groups will
be lower than is predicted by the semi-analytic model. 

Fortunately, the existence of an empirical color-density relation for
galaxies also suggests a purely empirical means for reproducing it in
mock catalogs, if our goal is simply to reproduce the features of the
data, without necessarily understanding the astrophysical processes
necessary to produce them in detail.  Just as we were able to populate
the N-body simulation with galaxy luminosities by making use of the
known relation between luminosity and halo mass, we can add galaxy
colors using the relation between color and local galaxy density.
This will be slightly more complicated than the approach we used
to add luminosities, since in that case we were able to posit a very
tight (indeed, zero-scatter) relation between mass and luminosity,
whereas the measured color-density relation in DEEP2 has an extremely
large scatter \citep{Cooper06a}, within which the galaxy colors take
their usual bi-modal distribution between red and blue objects.
Furthermore, there is also a strong correlation between galaxy color
and luminosity (and thus presumably also between color and
mass\footnote{Indeed, there is much debate in the literature at
  present regarding the relative importance of halo mass and local
  environment in determining galaxy color, with a large contingent
  arguing that mass is the more fundamental parameter
  \citep[e.g., ][]{Woo12}.  For our
  purposes, though, it is sufficent to note that these correlations
  exist in the data, whatever their underlying cause, and attempt to
  reproduce them in the mocks.})  via the well-known red and blue
sequences, and each of these correlations may be evolving with
redshift.

To ensure that we accurately capture the correlations between color,
luminosity, and galaxy environment, and the scatter and evolution in
these relations, we use a proceure that can be briefly outlined as
follows. First, we divide both the DEEP2 dataset and the mock catalog
into bins of redshift, luminosity, and local galaxy density.  Within
each bin, we select galaxies at random from DEEP2 (with replacement,
and after applying a weighting to account for the color-dependent
DEEP2 redshift-failure rate).  For each galaxy thus selected, we
assign its $k$-corrected $U-B$ color (as computed in W06) to one of
the galaxies from the same bin in the mock.  We explain each step in
this procedure in more detail below.  Our approach automatically
reproduces the measured color distribution from DEEP2 in each bin and
therefore also the evolving color-luminosity-density relation and its
scatter.  It is inspired by the
ADDGALS algorithm (Wechsler et al. in preparation) that was used to
produce mock catalogs for use with the Sloan Digital Sky Survey
cluster-finding efforts \citep[e.g.][]{Koester07b, Johnston07, Rozo07}
and several other purposes, including testing photometric redshifts
\citep{Gerdes10} and spectroscopic followup \citep{Cunha12}. Our
algorithm differs from ADDGALS, however, in its use of subhalo
abundance matching to assign luminosities and in various details of
the color-assignment algorithm---e.g., the use of broad-band colors,
rather than SEDs---that make it specific to DEEP2 mock-making.

\subsubsection{Choosing an environment measure}

We have not yet specified what metric we will use to measure local
galaxy density in this process.  \citet{Cooper05} tested a number of
different possible density measures in the YWC catalogs and found
that the distance to the $n$th-nearest
neighbor reproduced the underlying galaxy density field with the best
combination of fidelity and dynamic range.  The basic strategy is to compute the
projected distance $r_n$ to the $n$th-nearest neighbor within a window
of $\pm 1500$ \kms in the redshift direction.  This can then be converted
into a surface-density measure, 
\begin{equation}
  \Sigma_n = n/(\pi r_n^{2}).  
\label{eqn:sigman}
\end{equation}
Because the mean
density of a magnitude-limited galaxy sample falls with redshift, it
is also important to normalize the density to account for this; we can
do this by converting the density values to \emph{over}densities, 
\begin{equation}
\delta_n = \Sigma_n/\langle\Sigma_n(z)\rangle,
\label{eqn:deltan}
\end{equation} 
where the quantity in the denominator is the smoothed mean density
computed in rolling bins of redshift.  This measure is convenient for
our purposes since it is the one that was used to measure the
color-environment relation and its evolution in DEEP2
\citep{Cooper06a, Cooper07}.  In this work we will follow those
studies and measure local galaxy density in DEEP2 by computing the
third-nearest neighbor overdensity $\delta_3$, since $n=3$ was the
value that \citet{Cooper05} identified as giving the most accurate
measure of local density in DEEP2.

Because the DEEP2 targeting algorithm can only schedule $\sim 70\%$ of
appropriate targets for spectroscopy, and because the redshift-success
rate is $\sim 70\%$, the overall sampling rate of the
magnitude-limited pool of DEEP2 targets is $\la 50\%$.  When estimating
the local environment in the mocks for color assignment, we
are considering a complete, volume-limited sample, so 
the third-nearest-neighbor distance will probe the density field on
smaller scales in the mocks than it will in DEEP2.  It will be
preferable to choose a density estimator in the mocks that is
comparable to the $\delta_3$ values that would be measured after
accounting for all observational effects.  To find the appropriate
estimator, we computed various $n$th-nearest-neighbor overdensities
within an early version of these mock catalogs.  We then applied the
DEEP2 targeting algorithm and a simple redshift-failure rate (see the
following sections for details), and we computed the DEEP2 estimator
$\delta_3$ for the resulting ``observed'' catalog.  A comparison of
the various density measures showed that the projected
\emph{seventh}-nearest-neighbor overdensity in the volume-limited mock
$\delta_7$ was most tightly correlated with the as-observed
$\delta_3$ values.  Therefore we will use $\delta_7$ as our density
estimator for color assignment in the mocks.

A possible complication is that, even if $\delta_3$ and $\delta_7$ are
strongly correlated, their absolute values may not be directly
comparable.  It is not guaranteed that the correlation will lie along
the line of equality on a plot of $\delta_3$ versus $\delta_7$.  One
might suppose, for example, that the maximum overdensity in the
sparsely sampled DEEP2 catalog might have a lower numerical value than
the maximum in the volume-limited catalog.  Thus, simply binning the
data in $\delta_3$ and the mock in $\delta_7$ may not give comparable
samples in corresponding bins.  To deal with this, we divide each
catalog into quintiles of local density and assume that corresponding
quintiles of measured overdensity contain galaxies with comparable
physical local densities.  For example, the most overdense 20\% of
mock galaxies receive colors drawn from the most overdense 20\% of
DEEP2 systems, and so on.

A final concern for computing projected nearest-neighbor distances is edge
effects.  If a galaxy lies close enough to the edge of a field that
its true $n$th-nearest neighbor lies outside the field, then the
measured $\delta_n$ for that galaxy will be misestimated.  This
problem can be handled straightforwardly by imposing buffer
regions near the field edges.  In the DEEP2 data, we exclude from our
color-assignment algorithm all galaxies that lie  within 1 comoving
$h^{-1}$ Mpc from any edge or gap in the observed region.  This
ensures that only galaxies with accurately measured $\delta_3$ values
are used to assign colors.  In the mocks, we would like to assign
colors to all galaxies within the observed region, so we cannot
perform the same exclusion.  Instead, we construct our mock lightcones to
have an angular extent that is larger than a DEEP2 observed field by
0.2 degrees in either direction.  This creates an 0.1 degree buffer
all the way around the ``observed'' region of the mock, so that
galaxies in that area will have accurately measured values of
$\delta_7$.  

\subsubsection{Adding colors to the mock galaxies}
\label{sec:add_colors}

\begin{figure}
\includegraphics[width = 3.5in]{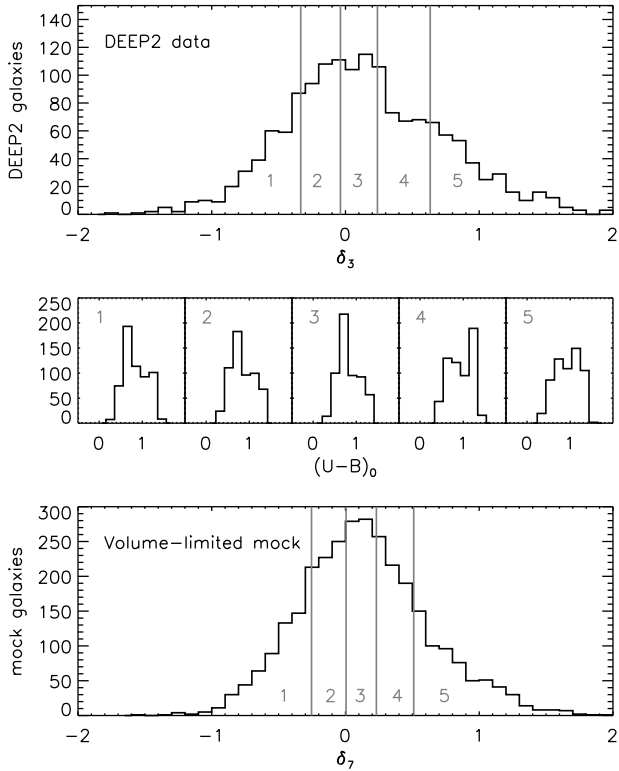}
\caption{Schematic representation of our basic algorithm for adding
  galaxy colors to the volume-limited mock catalogs.  Within a bin of
  redshift and absolute magnitude ($0.8\le z < 0.9$ and $-21\le M_B -
  5 \log h < -20$ is shown here), we divide the DEEP2 data into five
  bins of local density $\delta_3$, containing equal number of
  galaxies.  These bins are denoted by gray lines and numbers in the
  top panel.  We divide the mock galaxies similarly into quintiles of
  $\delta_7$ (bottom panel).  We then assign rest-frame $U-B$ colors
  to the mock galaxies in each density quintile according to the color
  distribution in each of the corresponding DEEP2 density quintiles,
  after weighting these distributions to account for the color-dependent
  redshift failure rate
  (middle panels).  }
\label{fig:color_mapping}
\end{figure}

Having specified our mapping between overdensity measured in DEEP2 and
in the complete, volume-limited mock, we can proceed to add colors to
the mock galaxies.
Figure~\ref{fig:color_mapping} gives a schematic picture of our
color-assignment technique.  Within a single bin of redshift and
$M_B$\footnote{A detail: our magnitude binning is performed after
  subtracting off the
  evolution of $M_B^\ast$ with scale factor $a$; that is, we construct
  bins of $L/L^\ast$.}, we divide the DEEP2 sample  into quintiles of
$\delta_3$ and the mock into quintiles of $\delta_7$.  We then draw
galaxy colors randomly from the galaxies in each DEEP2 quintile in
turn, and we assign them to the galaxies in the corresponding mock
quintile until all mock galaxies have been assigned colors.  Because
the DEEP2 redshift-success rate depends on color, before
performing the random draw, we also weight the data by the
incompleteness-correction weights computed in W06.  (In
the EGS field, we also divide out the color-dependent weighting that
was applied in the DEEP2 target-selection algorithm for this field, to
ensure that we have an unbiased color distribution.)

\begin{figure}
\includegraphics[width=3.0in]{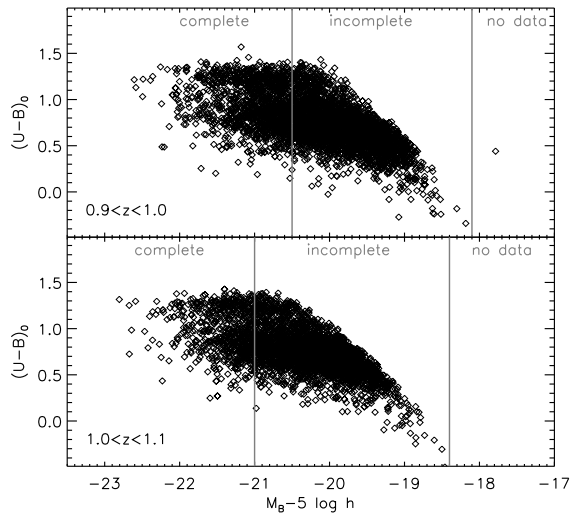}
\caption{An illustration of redshift and color-dependent
  incompleteness in DEEP2.  The panels show rest-frame color-magnitude
diagrams for two different redshift bins, as computed using the
$k$-corrections of W06.  The sharp, tilted cutoff at the
faint end corresponds to the DEEP2 $R=24.1$ magnitude limit at the
lower redshift limit of each bin.  Vertical gray lines in each panel
show the point at which the galaxy sample in each bin becomes
partially incomplete (\ie, some red galaxies start to drop below the
magnitude limit at the highest redshifts in the bin) and the point at
which all galaxies are lost from the sample.  (The very faint outlier
in the upper panel is an example of a galaxy with an incorrectly
identified redshift.  Such objects are rare and are
excluded by construction from our color-assignment algorithm.) }
\label{fig:incompleteness_illustration}
\end{figure}

This approach can be applied straightforwardly in bins of redshift and
$M_B$ that are completely sampled by the DEEP2 survey.  Within each
redshift bin, however, there is a color-dependent threshold
luminosity, below which DEEP2 constitutes a partially incomplete
sample and a minimum luminosity below which no DEEP2 galaxies are
observed.  Examples of these limits are shown in
Figure~\ref{fig:incompleteness_illustration}.  Because the DEEP2
$R$-band selection is perfomed in increasingly blue rest-frame bands
as redshift increases, the bias against faint red galaxies also
worsens.  To determine whether we are working in a partially
incomplete bin while assigning colors, in each bin we first perform a
test color assignment on a fraction of the mock galaxies.  We then
compute $R$-band apparent magnitudes for this sample, as described in
Section~\ref{sec:unkcorrect} below, and check for values that fall
below the DEEP2 magnitude limit, $R=24.1$.  Assigning colors to mock
galaxies in such bins will be more complicated than just described.
To make this process easier, we work through the redshift and
luminosity bins in increasing order of redshift and decreasing order
of luminosity, to ensure that, for each partially incomplete bin,
there is a nearby complete bin at lower redshift or brighter
luminosity that has already been populated with mock colors.

In partially incomplete bins, it will be necessary to exclude some
galaxies from the mock catalog by assigning them rest-frame colors that
will put them below the DEEP2 apparent magnitude limit.  We have no
direct information about the density distribution of galaxies below
the magnitude limit, but because the incompleteness depends on color,
it is likely that it also has some correlation with galaxy
environment.  Our task, then, is to determine \emph{(a)} how many
galaxies to exclude, \emph{(b)} with what distribution in local
density, and \emph{(c)} what colors to assign them so that they drop
out of the mock DEEP2 survey.  

The first step is relatively simple.  To get a rough estimate of the
number of mock galaxies to discard, we can simply compare the number
density, $n_\mathrm{mock}$, of mock galaxies in the
redshift-luminosity bin in question and compare it to the number
density of DEEP2 galaxies in the same bin, $n_\mathrm{DEEP2}$, after
weighting the DEEP2 galaxies with the W06
incompleteness-correction weights (which account for targeting
incompleteness and redshift failures, but not incompleteness owing to
the apparent magnitude limit).  However, one expects some
field-to-field variance in the mock owing to sample variance and shot
noise, and we would obviously like this variance to be reflected in
the mocks.  We can straightforwardly compute the impact of this
scatter on a particular mock lightcone by integrating the input
luminosity function (see Equation\ref{eqn:LF} and
Table~\ref{tab:LFparams}) over the luminosity range of the bin in
question and comparing the result to the actual number density in the
mock lightcone being considered.  This gives the size of the deviation
from the mean DEEP2 number density in that particular redshift bin of
that particular lightcone.  Then the target number density of observed
mock objects in that bin is given by
\begin{equation}
n_{\mathrm{targ}} = \frac{n_\mathrm{DEEP2} n_\mathrm{mock}}{\int_{L_\mathrm{min}}^{L_\mathrm{max}} \phi(L)dL},
\label{eqn:ntarg}
\end{equation}
where $L_\mathrm{min}$ and $L_\mathrm{max}$ are the upper and lower
limits of the luminosity bin being considered.  Having computed this
target number density, we can proceed to exclude galaxies from the
mock until $n_\mathrm{targ}$ is reached.

\begin{figure}
\includegraphics[width=3.5in]{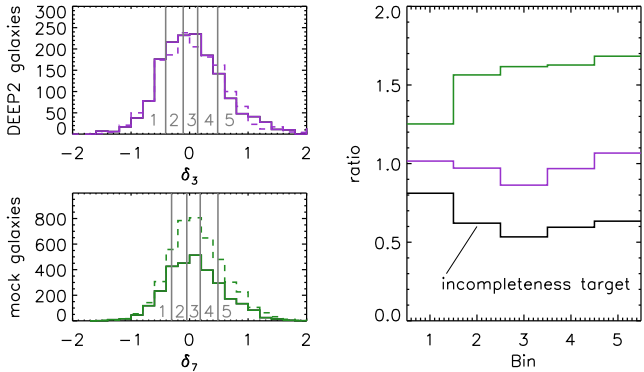}
\caption{A schematic diagram showing our technique (described in
  detail in the text) for rejecting
  galaxies in a local-density-dependent way from bins of luminosity
  and redshift  in which the DEEP2 sample is incomplete.}
\label{fig:density_rejection}
\end{figure}

Which galaxies should we exclude?  It is extremely likely that the
incompleteness in DEEP2 is a function of local environment, since it
depends strongly on color, and it is important that we ensure that the
overdensity distribution in the mocks is comparable to the one in the
data before we perform color assignment.  Since we cannot directly
measure the overdensity distribution of unobserved DEEP2 galaxies, we
will need to find some means of inferring it.  Our approach is to make
comparisons to the nearest complete bin at the same luminosity but
lower redshift.  The method is illustrated schematically in
Figure~\ref{fig:density_rejection}.  We start by dividing this lower
redshift bin into quintiles of overdensity, both in the data and in
the mock.  Then, \emph{holding the overdensity bins fixed}, we divide
our incomplete bin into the same five bins of overdensity, in both the
mock and the data.  This is shown in the left-hand panels of the
Figure, where the low and high-redshift bins are denoted by solid and
dashed lines, respectively .

We then compare the ratios of the total number counts in
each of the five overdensity bins (right-hand panel of the Figure).
It is to be expected that the distribution of overdensity values in
these bins will evolve in the mock catalog, owing to the evolution of
cosmic structure (green curve).  Because of density-dependent
incompleteness, we expect that the distribution will appear to evolve
differently in the data (violet curve).  By taking the ratio of the
relative change in the DEEP2 distribution to that in the mock (black curve), we can
compute the number density of galaxies that ought to remain in the $i$th bin
of overdensity after accounting for incompleteness:
\begin{equation}
\frac{n_\mathrm{targ}^i(z)}{n_\mathrm{mock}^{i}(z)}=
\frac{n_\mathrm{DEEP2}^i(z)}{n_\mathrm{DEEP2}^i(z_0)}
\left(\frac{n_\mathrm{mock}^i(z)}{n_\mathrm{mock}^i(z_0)}\right)^{-1}
\frac{n_\mathrm{targ}}{n_\mathrm{DEEP2}} 
\end{equation}
where $z_0$ is the redshift of the nearest complete bin at this
luminosity, and the final factor accounts for the expected sample
variance in the mock lightcone in question (see Equation~\ref{eqn:ntarg}).

Having set these targets for the number denisity of mock galaxies in
each bin of environment, we proceed to exclude galaxies from the
sample using an iterative process as follows.  Using the overdensity
binning that we established in the lower-redshift complete bin at
$z_0$, we assign provisional colors to the mock galaxies as described
above, using the lower-redshift complete DEEP2 sample as the input
sample.  We then convert the mock galaxies colors and absolute
magnitudes to apparent $R$-band magnitudes as described below.  Since
the bin we are considering is incomplete, some of these objects will
have $R>24.1$ and hence will drop out of the observed sample in the
mock.  For these galaxies, we retain the colors we just
assigned\footnote{If the number of galaxies with $R>24.1$ is larger
  than the number needed to reach $n_\mathrm{targ}^i$, then we retain
  the colors for only faintest ones up to this number.} and exclude
them from the sample; we erase the colors for the rest.  We then
repeat this procedure on the remaining galaxies until we have excluded
enough galaxies to reach $n_\mathrm{targ}^i$ in this bin.

For the remaining mock galaxies in this bin, we then assign colors
exactly as we did in the complete bins, by binning both the mock
galaxies and the DEEP2 galaxies at the \emph{current} redshift into
quintiles of overdensity, then drawing DEEP2 colors at random from
within each quintile to assign to the mock galaxies in the
corresponding density bin.  There is one small complication remaining,
however.  This bin has an incomplete region in color-magnitude space,
and it is possible that some of the color assignments in the remaining
mock galaxies will fall in this region.  To guard against this, we
test our color assignments by computing apparent $R$-band magnitudes
as described below.  For any galaxies with $R>24.1$, we erase their
colors and draw new ones, repeating this procedure until all the
remaining mock galaxies have colors that place them above the DEEP2
magnitude limit.  This completes the task of assigning colors to the
mock in partially incomplete bins of redshift and luminosity.  

For the faintest luminosity bins, where DEEP2 data is entirely
lacking, or where there is no complete bin at lower redshift, we
simply extrapolate from the color-density relation we used to populate
the next-brightest bin, with one modification.  Because galaxy
luminosity and color are correlated, we shift the colors that we
assign to the faint mock galaxies according to the mean relation,
treating red galaxies and blue galaxies separately.  First, we divide
the DEEP2 sample into red and blue galaxies according to the division
used by W06.  We then fit for the mean linear color-magnitude relation
of each of these samples seperately, finding that red galaxies redden
by 0.03 magnitudes per unit increase in absolute magnitude, while blue
galaxies redden with a slope of 0.1 magnitudes.  We use these
relations to shift the assigned colors blueward, according to the
difference in magnitude between the bright input galaxies and the
faint mock galaxies that are having colors assigned.  In this case,
the input galaxies we use to assign colors are not DEEP2 galaxies but
rather the galaxies in the next-brightest bin of the \emph{mock},
since the DEEP2 sample will be incomplete, but this has been accounted
for in the mock, as described above. (Thus, in this case, we use
$\delta_7$ as the density measure for both samples, but the algorithm
is otherwise identical.)  The color-density relation this produces is
unconstrained by data and thus likely to be incorrect in detail, but this is
irrelevant for any practical use of these mocks, and in any case the
situation cannot be improved without more data.

\subsubsection{Computing apparent magnitudes}
\label{sec:unkcorrect}

Once we have assigned rest-frame $B$-band luminosities and $U-B$
colors to the mock galaxies, we can convert these to apparent $R$-band
magnitudes $m_R$ by empirically inverting the $k$-correction procedure that
was used to compute the DEEP2 rest-frame values in W06.
The rest-frame values are related to the observed magnitude by 
\begin{equation}
m_R = M_B + D(z) - k,
\label{eqn:kcorrect}
\end{equation}
where $k$, the $k$-correction, is a function of $U-B$ and $z$;
and $D(z)$ is the distance modulus.  To obtain the appropriate
$k$-correction for our mock galaxies, we infer it from the values
computed for the DEEP2 sample as follows.

First, we divide the DEEP2 sample into narrow bins of width 0.01 in
redshift.  Within each of these bins, we fit a third-order polynomial
to the relation between the $k$-corrections and the rest-frame $U-B$
colors that were computed in W06.  As shown in the
Appendix of that paper, $k(U-B)$ is a smooth, nearly linear
function at fixed redshift, so our third-order polynomial fit will
capture the relation well, and extrapolating a small distance into
incomplete regions of color space will not be a problem.  We use the
fit thus obtained to compute $m_R$ for each mock galaxy according to
Equation~\ref{eqn:kcorrect}.  This technique is identical to the one
used in \citet{Gerke07a} to compute observed magnitudes in the YWC
mocks after assigning rest-frame colors in a manner similar to the one
described above.

\section{Simulating DEEP2 observations}
\label{sec:observation}
The mock DEEP2 lightcones we constructed in the previous section
should fully capture the impact of the DEEP2 $R=24.1$, but to fully
characterize the DEEP2 selection it will also be necessary
to account for the other observational effects present in the survey.
These effects comprise the DEIMOS slitmask-making algorithm that
schedules galaxies for spectroscopic observation; contamination by
foreground stars and background galaxies; and the magnitude and
color-dependent failure rate for obtaining
reliable redshifts.  We describe our
methods for modeling each of these effects in turn below.

\subsection{Mask-making: The DEEP2 targeting algorithm}

DEIMOS is a multiplexed slit spectrometer, in which a custom-made mask
is placed in the focal plane, with a slit cut in the position of each
object that has been scheduled for observation.  A number of practical
constraints control the number of objects that can be observed in this
manner.  First, since the light is dispersed along a particular
direction within the spectrograph, the slits must be oriented roughly
parallel to an axis perpendicular to the spectral direction (within
DEEP2 the slits are allowed to be tilted by up to 30 degrees from this
axis).  To avoid overlapping spectra on the detector, slits also may
not overlap along this spatial axis.  To allow for sufficient sampling
of night-sky emission for sky subtraction, the slits must be long
enough to allow empty regions of sky to be observed on either side of
the target galaxies; this limits slit lengths to be no shorter than
three arcseconds in DEEP2.

These limitations mean that galaxies in crowded regions on the sky
will be less likely than average to be assigned to a slit on a given
mask.  To mitigate the effects of slit-crowding, DEEP2 tiled the
survey region with slitmasks in an overlapping pattern, so that nearly
all galaxies have two or more opportunities to be assigned to a
slitmask\footnote{As discussed below, this increases to four chances
  in the EGS.}.  The DEEP2 slitmask creation algorithm is outlined in
~\citet{DGN04} and are described in detail by \cite{DEEP2}.

The mock DEEP2 lightcones we described above are constructed to have
an identical geometry to the three main DEEP2 fields, so the DEEP2
slitmask-making algorithm can be applied directly to the mocks to
select mock galaxies for spectroscopic ``observation'' (though the
mock lightcones must first be divided into three subcones,
corresponding to the three photometric pointings making up each DEEP2
field).  

The remaining complication is the pre-selection that is
performed in observed $BRI$ color-color space in the DEEP2 photometry.
Since we have only produced observed $R$ band apparent magnitudes, we
cannot directly apply this selection to the mock catalog.  The
color-color pre-selection is tuned to select a nearly complete set of
galaxies at $z\ge 0.75$, while excluding nearly all galaxies at $z\la
0.7$ \citep{DEEP2}, so it should be possible to approximate
it as a redshift-dependent selection weight.  DEEP2 galaxies in the
EGS were selected without this pre-selection, so we can use this field
to test the efficacy of the color selection and convert it into a
redshft selection.
The EGS sample confirms that the DEEP2 color cuts are a highly
efficient means of selecting high-redshift galaxies: only a few
percent of $z>0.75$ galaxies in the EGS are excluded by the cuts (and
many of these have large photometric errors), while a large
fraction of lower redshift objects are excluded.  Apart from a small
residual population of foreground galaxies (which we deal with below),
the DEEP2 color pre-selection can be accurately approximated as a
redshift-dependent selection probability as follows:
\begin{equation}
p_\mathrm{sel}(z) = \left(1+ e^{-43.7(z-0.725)}\right)^{-1}. 
\label{eqn:zselect}
\end{equation}

We apply this selection probability to the mock galaxies before
passing them through the DEEP2 slitmask making algorithm. We also
exclude all galaxies with $z>1.4$ in  this selection step, since the
[OII] $\lambda\lambda3727$ doublet leaves the DEEP2 spectral
window at this redshift, so the redshift success rate of DEEP2
galaxies drops dramatically at higher $z$.  We will account for these
higher redshift objects, as well as foreground objects and stars, below. This
accurately reproduces the redshift distribution of DEEP2 in the range
$0.7 \la z \le 1.4$, but it does not correctly account for the color
dependence of the selection near the redshift cutoff.  For this reason, the colors
of ``observed'' galaxies in the main (non-EGS) DEEP2 mock catalogs
presented here should be treated with caution at $z<0.75$.  

\subsubsection{Foreground and background objects}

The redshift selection function described above accurately
captures the DEEP2 selection in the primary redshift range $0.75 \le
z \le 1.4$.  However, it does not account for the small but
significant fraction of DEEP2 targets that do not lie in this range.
These include objects in three categories: foreground ($z<0.75$)
galaxies that nevertheless pass the DEEP2 color pre-selection,
foreground stars that have been misclassified as galaxies in the DEEP2
photometry, and background objects at $z>1.4$.   Since objects in each
of these categories use up slit length that could otherwise be
assigned to galaxies in the primary DEEP2 range, it is important to
account for their effects on the DEEP2 slitmask-making algorithm.
Since each of these samples is highly incomplete with complicated
selection, it is nearly impossible to accurately reproduce the
properties (\eg, colors) of the objects in each set.  Since such
objects are not typically the focus of scientific analysis,
and since we only wish to include their impact on the spectroscopic
targeting, it is reasonable to include them in a random way
as described below.

The sample of DEEP2 targets with reliable redshifts in the three
non-EGS fields consists of roughly $1\%$ stars and $4\%$ galaxies at
$z<0.75$. Follow-up of the DEEP2 redshift failures with UV
spectroscopy shows that an additional $\sim 15\%$ of DEEP2 targets lie
at $z>1.4$ (C. Steidel, private communication).  To contaminate the
sample with stars before applying the maskmaking algorithm, we draw
$R$-band apparent magnitude values from a Gaussian distribution with
mean 21.9 and dispersion of 1.5 magnitudes.  We draw from this
distribution and create potential spectroscopic targets (``stars'') at
redshift zero in sufficient numbers to constitute $1\%$ of the
targets.  To produce foreground and background galaxies, we select
galaxies at random from the populations at $z<0.75$ and $z>1.4$ in
appropriate proportions to match what is found in DEEP2, and we add
these to the pool of potential spectroscopic targets.  Once these
contaminants are added to the mock catalog, we execute the DEEP2
slitmask making procedure on this sample, producing a list of objects
that would be scheduled for DEIMOS observation in a real survey.  This
constitutes our mock DEEP2 spectroscopic sample.

\subsection{Simulating redshift failure}

Approximately $30\%$ of DEEP2 spectroscopic targets do not yield good
redshifts \citep{DEEP2}.  As we mentioned above, roughly
half of these lie in the so-called redshift desert at $z>1.4$, where
no strong galactic spectral features fall in the optical wavelength
range.  The remaining $15\%$ fail to yield redshifts for a variety of
reasons, most of which lead the observed spectrum to have a low signal
to noise ratio.  This redshift
failure rate has some dependence on apparent magnitude and color: for
example, faint red galaxies combine low flux with weak spectral
absorption features, making them particularly susceptible to redshift
failure.  If we wish our mock catalogs to accurately model the
selection function of galaxies with DEEP2 redshifts, it is important
that we include color and magnitude dependence of redshift failure in
our catalogs.  

W06 developed a weighting scheme that accounts for color and
magnitude-dependent incompleteness in DEEP2 by binning the
spectroscopic targets in observed color-color-magnitude space and
comparing the number of successful redshifts in each bin to the total
number of photometric objects (making some justified assumptions about
the redshift distribution of the spectroscopic targets that failed to
yield a redshift).  These weights can be straightforwardly inverted to
yield an estimate of the redshift failure probability for a galaxy in
a particular region of photometric space.  To apply this probability
to the mocks, when we assign rest-frame $(U-B)$ colors to the mock
galaxies by randomly drawing DEEP2 objects from within a bin of
density, we also assign each DEEP2 galaxy's incompleteness weight
$w_\mathrm{corr}$ to the mock galaxy that receives its color.

We use the ``optimal'' set of weights from W06 for this purpose, with
two small modifications.  The weights account for incompleteness owing
both to redshift failure and for targeting incompleteness (\ie, the
failure to schedule all suitable photometric objects for
spectroscopy).  Since we have already accounted for the latter type of
incompleteness in the mock by running it through the DEEP2
slitmask-making algorithm, we must correct for this to avoid
double-counting the targeting incompleteness.  The
slitmask-making algorithm does not depend on color or apparent
magnitude, so the correction is simple: we need only multiply each weight
by the fraction of potential targets that are scheduled for
spectroscopic observation.  (For galaxies in the EGS, it is also
necessary to multiply out the redshift-dependent selection weighting
described in the next section, which is straightforward.)  In
addition, there is a sharp drop in redshift success for the
faintest 0.5 magnitudes of the DEEP2 sample (i.e., galaxies with
$23.6\la R<24.1$).  This is accounted for in the weighting scheme, but
there is no guarantee that a mock galaxy will have exactly the same
apparent $R$ magnitude as the DEEP2 galaxy we used to assign its
weight.  Hence, if either the source DEEP2 galaxy or the target mock
galaxy has $R>23.6$, we apply a linear correction  to the mock galaxy's
weight with a slope of 0.13 per unit magnitude difference between the
target and source galaxy (applying a minimum of $R=23.6$ on both
galaxy magnitudes when computing the difference).

Each galaxy in the mock can then be assigned a probability of yielding
a successful redshift, $p_z = w_\mathrm{corr}^{-1}$, where
$w_\mathrm{corr}$ is the W06 incompleteness weight, corrected for the
targeting efficiency.  We then use these probabilities to select
stochastically a subsample of the spectroscopic targets that we
declare to have ``failed'' redshifts, until the fraction of redshift
failures (including both these stochastic failures and background
objects at $z>1.4$) matches the value in DEEP2.  We will refer to the
remaining targets as the ``observed'' mock galaxies. With this step, we have
fully accounted for the color and magnitude-dependent DEEP2 selection
effects in the mock catalogs, to the extent that it is possible to do so.

\subsection{The Extended Groth Strip}

In the EGS, spectroscopic targets are
drawn from all regions of color-color space; no pre-selection is
applied to exclude low-redshift galaxies.  (However, as described
below, the selection is \emph{weighted} in color-color space, to stop
the sample from being dominated by the more numerous low-$z$
galaxies.)  A sample that is magnitude-limited at $R=24.1$ and extends
to low redshift will include many nearby galaxies with very
low-luminosity---and hence low mass.  According to the W06
$k$-corrections, galaxies in the EGS have absolute magnitudes as low
as $M_B \sim -11$, similar to the Leo dwarfs; the abundance-matching
prescription we developed in Section~\ref{sec:abundancematch} would
place such galaxies in halos with masses below $10^{10} M_\odot$.  To
include such low-mass halos in N-body models with cosmologically
interesting volumes requires very high-resolution simulations, such as
Bolshoi, which have only been computationally feasible within the last
few years.  The use of such simulations is one of the primary
improvements of this work over the earlier mock catalogs of YWC04; it
allows us to construct the first realistic mock catalog that includes
the full range of redshift and luminosity probed by the EGS
spectroscopic dataset.

When constructing the DEEP2 mock catalogs, we made use of data
covering the full redshift range of EGS, $0<z<1.4$, making use only of
data from the EGS at $z<0.75$, where the primary DEEP2 spectroscopic
sample is highly incomplete.  Thus, our mock galaxies cover the full
range of photometric and redshift space that is probed by DEEP2
spectroscopy in the EGS.  However, there remain 
differences between the EGS dataset and the rest of DEEP2 that we must
account for in constructing an EGS mock.  The first is the field
geometry and orientation.  The three non-EGS DEEP2 fields are
$0.5^\circ \times 2^\circ$ rectangular fields with their long axes
oriented along lines of constant declination.  The EGS, by contrast,
is half as wide, at $0.25^\circ \times 2^\circ$, and is oriented
perpendicular to the ecliptic, to allow easier access by space-based
instruments.  To account for these differences when construcing EGS
mocks, we rotate the coordinate system of each of our mock lightcones
to match the orientation of EGS and then excise the central
$0.25^\circ$ strip to use in EGS-like maskmaking.  It is important to
note that this approach means that our EGS lightcones are \emph{not}
independent from our main DEEP2 mocks; instead, each EGS cone
corresponds exactly to one of the primary DEEP2 mocks.

The remaining differences between the EGS and the rest of DEEP2
involve the details of spectroscopic target selection.  As we
mentioned above, spectroscopic targets in the EGS are chosen from all
regions of photometric color space, in contrast to the main DEEP2
sample, where a color pre-selection is applied.  However, since a
simple, magnitude-limited survey will always be dominated by the
numerous fainter objects at low redshift, and since DEEP2's primary
goal is to explore the $z\sim 1$ universe, the EGS spectroscopic
target selection is weighted in color space, so as to
preferentially select high-redshift objects.  The particulars of this
weighting are discussed in detail by \cite{DEEP2}.  We briefly
summarize it here.  Galaxies in the EGS photometric sample are divided
according to the same color pre-selection as is used in the primary
DEEP2 fields.  All galaxies that pass this selection cut (\ie, the
ones that would be observed in the rest of DEEP2), along with all
galaxies brighter than $R=21.5$, receive a uniform selection
probability $p_0$.  For fainter galaxies that fail the selection cut
in color space, the probability of spectroscopic observation is a
function of flux, chosen so that roughly equal numbers of galaxies are
observed above and below $z=0.75$.  This selection weighting is
described in detail in \cite{DEEP2}.

We apply this same probabilistic selection weighting in the
spectroscopic targeting algorithm we use in constructing the EGS
mocks, with the difference that the selection in color space is
replaced with the selection in redshift given in
Equation~\ref{eqn:zselect}.  The EGS targeting algorithm also differs
from the primary DEEP2 selection algorithm in its approach to tiling
the observational area with slitmasks.  Whereas the main DEEP2 survey
strategy uses overlapping masks with a two-pass approach, the EGS
targeting algorithm gives most galaxies \emph{four} chances to be
scheduled for observation, with one half of the masks oriented
perpendicular to the other half, to further reduce crowding effects.
We use an identical algorithm to ``schedule'' our mock EGS galaxies
for spectroscopic observation.  Then, after accounting for redshift
failures as described in the previous
section, we have a simulated DEEP2 EGS spectroscopic catalog that
accounts for all major color and luminosity-dependent selection
effects.

\section{Comparisons to data}
\label{sec:compare}

Our aim is to produce mock catalogs by populating N-body models to
reproduce various properties of the DEEP2 sample as accurately as
possible.  These include the luminosity function, luminosity-dependent
two-point correlation function, color-magnitude diagram, and
color-luminosity-environment relation---as well as the evolution of
each of these with redshift.  As discussed above, there is good reason
to believe that reproducing these properties simultaneously will also
correctly reproduce the underlying relation between luminosity, color,
and dark-matter halo mass, thereby allowing the DEEP2 halo-mass
selection function to be inferred.

In order to gauge our success in reproducing DEEP2 properties, it is
obviously crucial to make direct comparisons to the data.  This will
also help to reveal any limitations of our mocks.  We have already
made comparisons above between our mock catalogs and the DEEP2
luminosity function (which our algorithm reproduces by construction;
Figure~\ref{fig:LF}) and the projected two-point correlation function
(Figure~\ref{fig:wprp_comparison}) for different luminosity
thresholds. Because of the evolving color-dependent selection cut
imposed by the DEEP2 magnitude limit, it is also important to test the
color, environment, and redshift properties of the mocks against the
data as well.  We make these comparisons in the remainder of this
section.  Throughout, we will use the Bolshoi mock catalogs as our
basis for comparison to the data, since its background cosmology is in
the best agreement with current constraints, but it will also be
interesting to investigate the impact that different cosmological
assumptions have in our three sets of mocks.  We will conclude the
section by briefly considering this issue.

\subsection{The color-magnitude diagram}

\begin{figure}
\includegraphics[width=2.75in, angle=90]{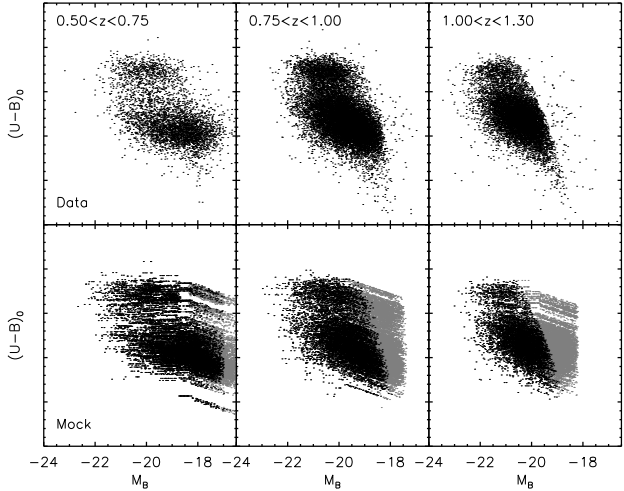}
\caption{Comparison of the color-magnitude diagrams, in bins of
  redshift, for DEEP2 (upper row) and the mock catalogs presented here
  (lower rows).  For the mock panels, black points show galaxies that
  were scheduled for observation and received a good redshift, while
  unobserved galaxies and redshift failures are shown in gray.  The
  sources of the striping patterns visible in the mock panels are
  explained in the text.}
\label{fig:colmag_compare}
\end{figure}

The first requirement for reproducing the DEEP2 color-dependent
selection effects in our mocks is that we accurately reproduce the
DEEP2 color distribution, its correlation with luminosity, and its
redshift evolution.  Figure~\ref{fig:colmag_compare}  shows the
rest-frame color-magnitude diagram for DEEP2 and the mock catalogs.
For the mocks, both observed galaxies (black points) and unobserved
galaxies (gray points) are shown.  The qualitative features of the
DEEP2 color-magnitude distribution are broadly reproduced in the
mocks: distinct red and blue sequences are evident at all redshifts,
with the correct loci in color-magnitude space and the correct
color-magnitude correlations for observed galaxies; and the
color-dependent selection cut in the mocks matches the one in the
data (apart from a small number faint outliers in the data, which
arise from incorrect redshift assignments). 
 
Clear artifacts of our color-assignment process are evident in the
mock diagrams, however, in the form of striping patterns.   Two
different kinds of stripe are apparent: horizontal stripes at bright
magnitudes (especially at low redshift), and diagonal stripes for
fainter objects.    The latter stripes result from our method for
assigning colors in luminosity bins where DEEP2 is incomplete,
described in Section~\ref{sec:add_colors}.  In these bins, we have
used the colors from brighter galaxies, shifted according to the mean
color-magnitude relations for red and blue galaxies in DEEP2; the
diagonal stripes for faint galaxies correspond to these relations.
Horizontal striping occurs when the number of galaxies in a given
DEEP2 luminosity-redshift bin is comparable to or smaller than the
number of galaxies in the corresponding mock bin.  This occurs
especially at low redshift, since the EGS field has only half the area
of the rest of the DEEP2 fields.  For this reason, the low-redshift
mock color-magnitude diagrams show a substantial striping and are
extremely noisy.  

We have not shown the $z<0.5$ diagram in
Figure~\ref{fig:colmag_compare}, since this source of noise is so
great there  as to make comparison to the data uninformative.  Since
the comparison in the $0.5<z\le 0.75$ bin is not especially helpful 
either, for the remainder of this section we will focus on comparing
the mocks with the high-redshift DEEP2 fields, excluding the EGS.  If
our techniques are successful at reproducing the high-$z$ DEEP2
properties, then they should be equally successful in the EGS, within
the confines of the limited data available in that region.

\begin{figure}
\includegraphics[width=3.5in]{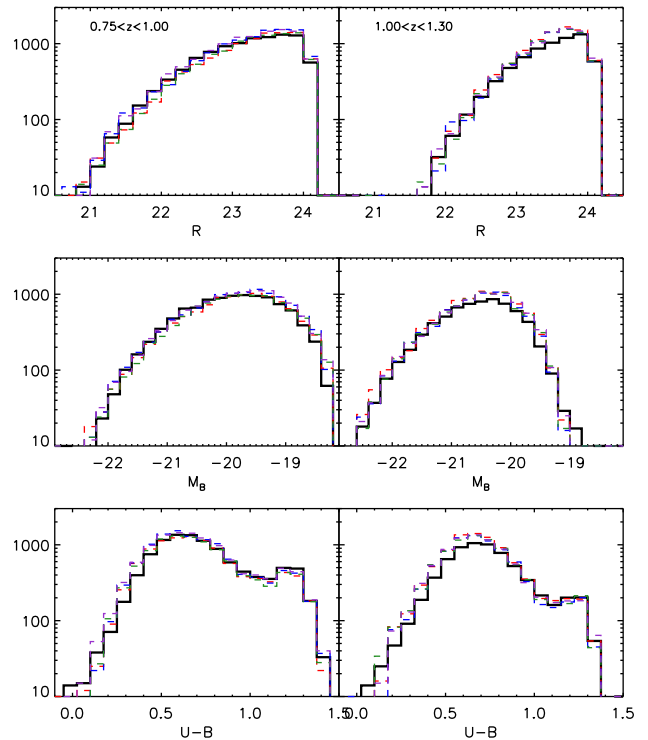}
\caption{Comparison of the apparent and absolute magnitude and
  rest-frame color distributions for the DEEP2 sample in 
the three high-redshift fields (black solid curves) and for four different
mock realizations of identical volume to those fields (colored dashed curves).}
\label{fig:colmaghist_compare}
\end{figure}

To allow a more quantitative comparison of the color and magnitude
distributions, we plot histograms of apparent magnitude, absolute
magnitude, and rest-frame color, over two different redshift ranges, in
Figure~\ref{fig:colmaghist_compare}.  To make direct comparisons
between the data and the mocks, we choose a subset of the 40 Bolshoi
light cones with equal area to that covered by the high-redshift DEEP2
fields; this constitutes a single mock realization of the DEEP2
sruvey.  The figure shows histograms for the DEEP2 data (black solid
lines) and for the observed galaxies in four different mock
realizations (colored dashed lines); the four realizations give a
sense of the scatter in these distributions.  The agreement is quite
good in general, although there is a small but systematic
overabundance of faint, blue galaxies, especially at higher
redshifts.

There are two explanations for this.  First, we have used the best fit
to the evolving DEEP2 luminosity-function paramters to populate our
mocks, rather than using the best fitting luminosity function at each
redshift.  This means that the mock luminosity function may lie above
or below the DEEP2 one in a given redshift bin.  As shown in
Fig~\ref{fig:LF}, the input luminosity function (dotted curve) lies
above the best-fitting DEEP2 luminosity function in the $1.0<z<1.2$
bin.  This explains why the high-$z$ absolute magnitude distributions
for the mocks lie systematically above the DEEP2 distribution, for
example.  However, the excess appears to be more significant at
fainter magnitudes, and this faint-end excess persists in the
low-redshift bin, where the input luminosity function matches the
DEEP2 best fit more closely.  This is likely explained by inaccuracy
the faint-end slope of the input luminosity function.  W06 held this
parameter fixed at a value of 1.30, since DEEP2 does not extend to
faint enough magnitudes for a robust fit, and we have adopted this
value here (Table~\ref{tab:LFparams}). Inspection of Figure 7 of W07
suggests that this fixed faint-end slope may slightly overestimate the
faint-galaxy abundance in DEEP2 at $z> 0.6$,  so it is not too
surprising that the mocks show a slight excess of faint objects.  This
excess translates into the systematic overabundance of blue galaxies
that is also apparent in Figure~\ref{fig:colmaghist_compare}, since
all of the faintest galaxies in DEEP2 are blue galaxies.

\subsection{The observed redshift distribution}

\begin{figure}
\includegraphics[width=2.75in, angle=90]{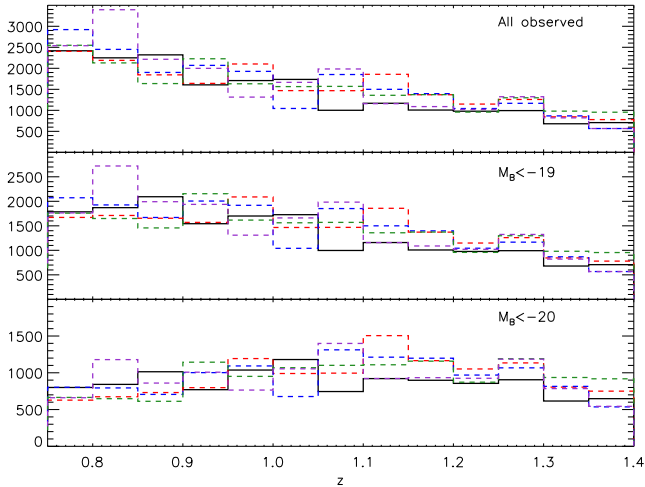}
\caption{Comparison of the redshift histograms for the DEEP2 sample in
  the three high-redshift fields (black solid curves) and for four
  different mock realizations of identical volume to those fields
  (colored dashed curves).  The mock realizations and their color
  coding are the same as was used in
  Fig~\ref{fig:colmaghist_compare}.  The panels show all observed
  galaxies (top), galaxies brighter than $M_B=-19$ (middle), and
  galaxies brighter than $M_B=-20$ (bottom).}
\label{fig:zhist_compare}
\end{figure}

Another test of the color and magnitude distribution in the mocks is
to compare the redshift distribution in the mocks to the DEEP2
distribution.  Reproducing the redshift distribution is not trivial:
since the DEEP2 magnitude limit is a strong function of redshift and
rest-frame color, an accurate redshift distribution requires accuracy
in the overall galaxy abundance, color distribution, color-luminosity
relation, and the evolution of each of these with redshift.  This
means that the redshift distribution is a strong test of our
mock-making techniques.  Figure~\ref{fig:zhist_compare} shows the
redshift distribution for DEEP2 and for the same four mock
realizations used above, for all observed galaxies, and for two
different thresholds in luminosity.  The agreement is quite good,
considering the scatter between the different mock realizations,
except for a small but systematic overabundance of galaxies over the
range $1.05<z<1.3$.  As explained in the previous section, this excess
can be explained by the difference between our input luminosity
function and the true best-fit DEEP2 luminosity function over this
redshift range.  It would be possible to improve the agreement here by
using the true best-fit DEEP2 luminosity function in each redshift
bin, but this would be at the expense of a smoothly evolving
luminosity function and would likely create unphysical and undesirable
jumps in the mock galaxy abundance at particular redshifts.  We thus
consider the current distribution to be the best that can be achieved
with the available data.

\subsection{The relation between color and environment}

\begin{figure*}
\centering
\includegraphics[width=3.in]{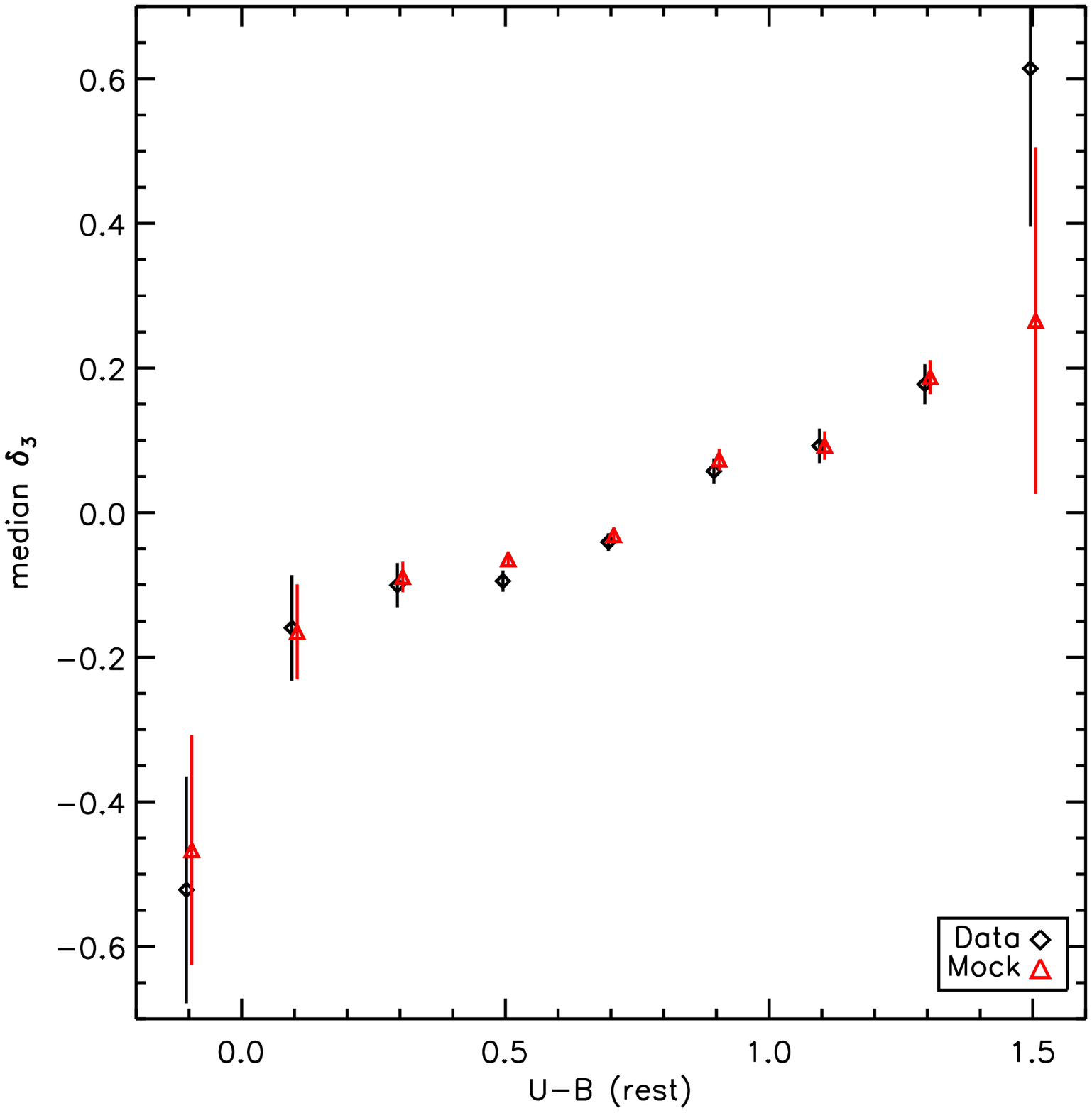}
\includegraphics[width=3.in]{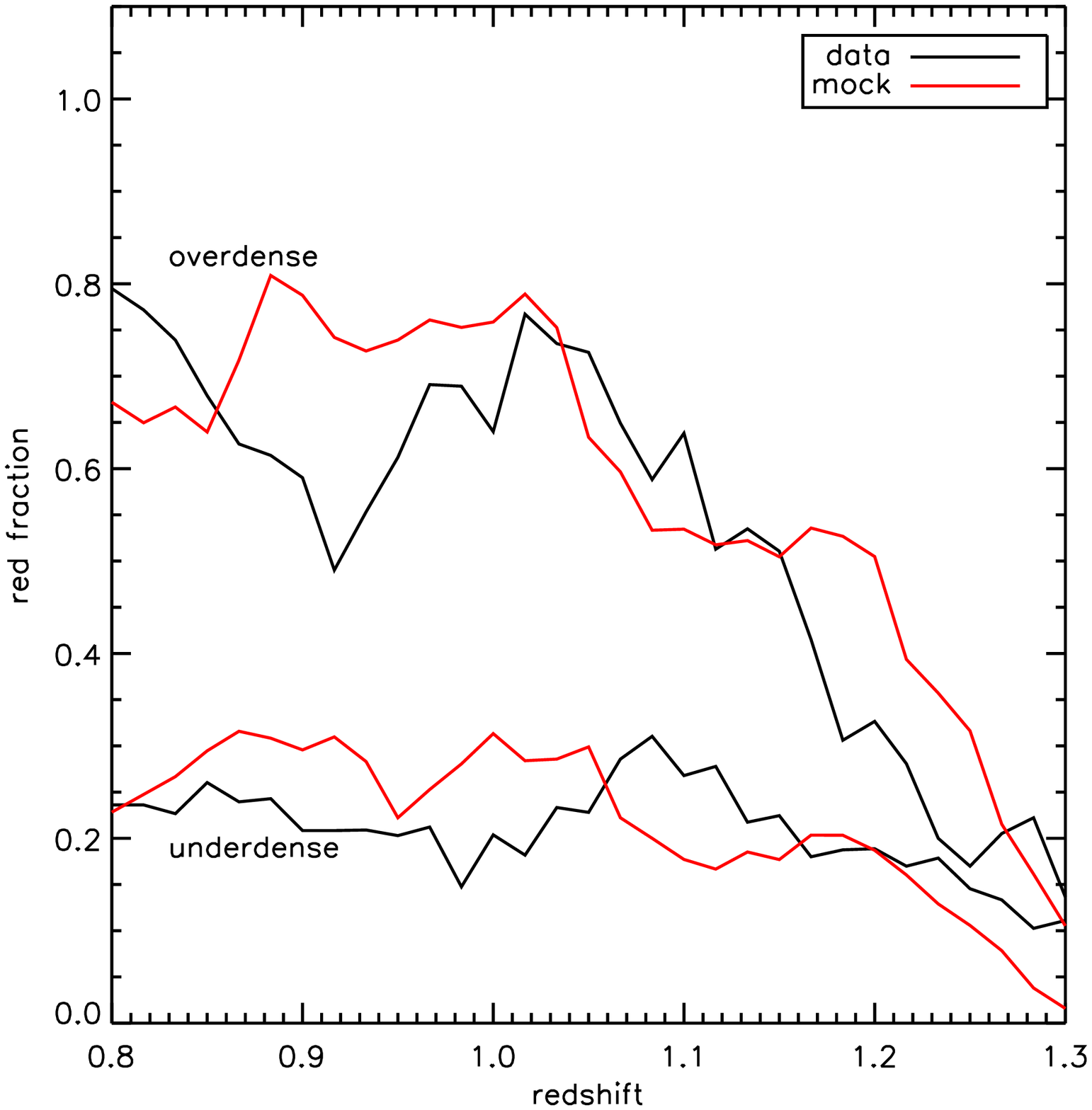}
\caption{Comparison of the color-environment relation, and its
  redshift evolution, in DEEP2 and the mock catalogs.  \emph{Left:}
  The median value of the local overdensity parameter $\delta_3$ is
  shown in bins of rest-frame color, for the DEEP2 data (black points)
  and the mock catalogs (red points).  Error bars show the standard
  error, and points have been offset slightly in the horizontal
  direction for clarity.  Here $\delta_3$ in the mocks has been
  computed using only observed mock galaxies, as for DEEP2.
  \emph{Right:} A comparison of the evolution of the color-density
  relation in the data and the mocks.  Curves labeled \emph{overdense}
  and \emph{underdense} show the fraction of red galaxies for the most
  overdense and most underdense $20\%$ of galaxies, respectively,
  where red galaxies are defined according to the red--blue division
  in W06.  Black curves show the red fractions in DEEP2, and red
  curves show the values from the mock catalogs.  Error bars on these
  curves are similar to the ones shown in Figure 6 of \citet{Cooper07}
  and are much smaller than the difference seen between the samples at
  low $z$; they are suppressed here for the sake of readability.}
\label{fig:environ_color_compare}
\end{figure*}

As we discussed in Section~\ref{sec:envmatch}, if we are going to
reliably infer the DEEP2 halo selection function, it is essential that
our mock catalog accurately reproduce the relation between galaxy
color and environment, as well as the redshift evolution of this
relationship.  We have attempted to ensure this by assigning colors to
our mock galaxies as a function of the local galaxy overdensity.
Our method for doing this relied on comparing density values measured
in the magnitude-limited DEEP2 sample to values measured in the
volume-limited mock, and so we matched the colors in bins of
\emph{relative} overdensity, using the $\delta_3$ parameter in the
data and $\delta_7$ in the mock.  There is no obvious guarantee that this will
translate to a mock that matches the \emph{absolute} color-density
relation from DEEP2.  It is important that we compare the
color-environment relation in the observed mock with the DEEP2 data,
to see how well the technique has worked.

This comparison is shown in Figure~\ref{fig:environ_color_compare}.
The left panel shows the median third-nearest-neighbor galaxy
overdensity $\delta_3$, in bins of rest-frame $U-B$ color, over the
redshift range $0.8<z<1.0$.  The black diamond points constitute an
update to the results of \citet{Cooper06a} and show the DEEP2
color-density relation for the final DEEP2 galaxy sample.  The red
triangles show the same quantities for the mock catalogs, where
$\delta_3$ has been computed using only the observed mock galaxies
(i.e., those that were targeted for observation and received a valid
redshift).  The agreement between the two samples is remarkable,
especially given the large number of selection effects
(apparent-magnitude limit, spectroscopic targeting, and
color-dependent redshift failure) that were applied to the mock
catalog after the initial color assignment was done.  This figure 
constitutes a strong confirmation that our method for adding colors in
quintiles of environment accurately reconstructs the DEEP2
color-environment relation.  Although this
figure shows only median values of $\delta_3$
over a narrow redshift range, we have confirmed that the full
$\delta_3$ distributions have similar shapes and that similar results
obtain for different redshift ranges.

Because the DEEP2 color-environment relation also evolves strongly
with redshift \citep{Cooper07}, it is also important that we capture
this evolution in the mocks as well.  The right-hand panel of the
igure compares this evolution in the data and the mocks.  The black
curves are an update to Figure~6 of \citet{Cooper07}: they show the
fraction of galaxies that are red in the most overdense and most
underdense $20\%$ of the DEEP2 sample, in sliding bins of redshift
with width 0.1 (where red galaxies are defined by splitting the red
and blue sequences as in W06).  The red fraction of underdense
galaxies is roughly constant with $z$, while the overdense red
fraction evolves strongly, especially at $z>1$, so that the two
populations are nearly indistinguishable in terms of red fraction at
$z=1.3$.   The red curves in the figure show the same quanities for
the mocks; the same qualitative behavior is evident, and the agreement
between the mocks and the data is quite good over the full redshift
range considered.  

The excellent agreement between the mocks and the DEEP2 data in terms
of redshift, color, and color-density relation suggests that these
mocks will be useful for exploring the impacts of DEEP2's color,
redshift, and density selection functions on the selection of dark
matter halos.  The Appendix gives an example of how the mocks might be
used for this purpose.

\subsection{The Halo-occupation distribution and the impact of cosmology}
Before we conclude, it will be interesting to explore the importance
of the background cosmology on the accuracy of the mock catalog.
Recall that we have constructed DEEP2 mocks for theree different
simulation boxes with three different cosmologies (listed in
Table~\ref{tab:cones}).  Throughout all of the above comparisons, we
have considered only the Bolshoi simulation, since its background
cosmology is the most consistent with current cosmological
constraints, and we have found that the Bolshoi mocks will be useful for
probing the DEEP2 mass selection function at high halo masses. We
would like to know the other mocks are likely to be similarly useful
or not.

\begin{figure}
\includegraphics[width=3.5in]{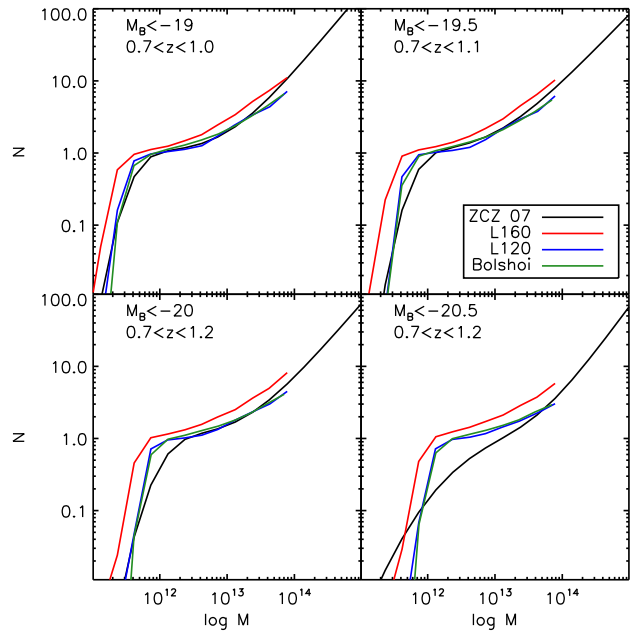}
\caption{Halo occupation distributions for the three different mock
  cosmologies, with different luminosity thresholds, compared to the
  analagous 
  HODs that \citet{ZCZ07} inferred for DEEP2.  A clear dependence of
  the HOD on background cosmology (explained in detail in the text) is
evident.  The L120 and Bolshoi mocks agree reasonably well with the
data in all but the brightest bin (where the inferred DEEP2 HOD, as
discussed in the text, is highly uncertain and problematic). }
\label{fig:hod_compare}
\end{figure}

The most direct way to test this is to consider the HOD.  Since the
halo-occupation probability will feed directly into the inferred halo
selection function, if this quantity depends strongly on cosmology it
will be especially important to make sure our simulation box has a
background cosmology that is consistent with existing constraints.  If
the dependence is weak, then this is less critical.  We can compute
the HOD for each of our mocks by dividing the dark matter halos into
bins of mass and counting the mock galaxies in each halo above a given
luminosity threshold, and then averaging these counts over the mass
bin.    It will also be interesting to compare these mock HODs to the
HOD that \citet{ZCZ07} (hereafter ZCZ07) inferred from the DEEP2 clustering
measurements.  This will constitute another comparison of the mocks to
the data, albeit a model-dependent one, since the inferred DEEP2 HOD
makes assumptions about the background cosmology.

Figure~\ref{fig:hod_compare} shows the comparison for the same four
luminosity thresholds that were used in ZCZ07.  There is a clear
difference between the mock from the L160 simulation box and the other
two boxes, with subtler differences between the L120 and Bolshoi
mocks.  The reason for these differences is easy to explain.  The
luminosity-assignment algorithm inserts a fixed luminosity function
$\phi(L)$, regardless of the underlying halo mass function.  Thus, if
the overall mass function has a lower normalization, we must place
more galaxies in halos of a given mass to reproduce the desired
luminosity function.  Hence, a simulation with a lower normalization
for its halo mass function (e.g., lower $\Omega_M$ or $\sigma_8$) will
necessarily have a higher mean halo occupation number $\langle
N\rangle(M)$ at fixed $\phi(L)$.  Conversely, because a galaxy of a
given luminosity would reside in a lower-mass halo, clustering at
fixed luminosity would be reduced. The opposite effect would obtain in
a simulation with a higher normalization.

Thus, the L160 mock, with its
relatively low value of $\sigma_8$ has a significantly higher HOD at
all masses than the other mocks.  The differences between the L120 and
Bolshoi HODs are smaller because their cosmologies (see
Table~\ref{tab:cones}) lie near a degenerate curve in
$\Omega_M-\sigma_8$ space; that is, there is little difference betwen
their mass functions.  It is worth noting, though, that the Bolshoi
mock HOD lies systematically above the L120 one in the mass range
$10^{12} \la M \la 10^{13}$, which is an important transition region
between the central-dominated and satellite-dominated parts of the
HOD.  Because there is a clear dependence of the HOD on cosmology in
these mocks, we recommend that the Bolshoi mocks be used for nearly
all purposes, since the others may give inaccurate selection
functions.  However, we also make the L120 and L160 mocks available to
the public, \eg, for the purposes of testing cosmology dependence.

The Bolshoi and L120 mocks also agree reasonably well with the HODs
inferred from DEEP2, for all but the brightest galaxy sample.  Black
curves show the HOD models from ZCZ07 that best fit\footnote{The one
  exception to this is the lowest luminosity bin.  In that bin, the
  formal best fit lies in a region of parameter space that ZCZ07
  excluded on physical grounds, so in this panel we show a curve that
  lies within the parameter uncertainties and that approximately
  reproduces the curve in Figure 1 of ZCZ07.} the DEEP2 clustering
measurements from \citet{Coil06b}. This agreement is not surprising,
since ZCZ07 assumed a background cosmology identical to the one used
in the L120 simulation.  For the three lower luminosity thresholds,
the Bolshoi and L120 HODs lie slightly below the DEEP2 measurement at
high mass, and they cut off somewhat less steeply at low mass.  There
is a strong degeneracy in the ZCZ07 HOD model, however, such that
increasing the sharpness of the low-mass cutoff can be offset by
lowering the high-mass slope while keeping the predicted
autocorrelation function roughly constant, so an even better agreement
is likely possible within the model uncertainties.  Since our zero-scatter
model effectively requires a sharp low-mass cutoff, it is
also possible that a model with somewhat higher scatter would be in
closer agreement with the ZCZ07 results.  

In the panel with the brightest luminosity threshold, the agreement
with the inferred DEEP2 HOD is quite poor.  However, the ZCZ07 fit in
this bin is poorly constrained and depends strongly on the
corrections that were applied to account for small-scale effects from
crowding on the DEEP2 slitmasks.  Furthermore, the ZCZ07 curve shown
there is unphysical (as ZCZ07 discuss), since it formally implies that
halos with $M=2\times10^{11}$ contain more galaxies with $M_B <-20.5$
on average than they do galaxies with $M_B < -20$.  We thus do not
find the disagreement at bright magnitudes to be a particular cause
for concern.

\section{Summary and Conclusions}
\label{sec:conclusions}

We have constructed three sets of mock catalogs for the DEEP2 Galaxy
Redshift Survey, derived from three different $N$-body simulations
with different background cosmologies.  In doing this, we have striven
to reproduce, as accurately as possible, the DEEP2 HOD, as well as all
of the important galaxy properties that might have an impact on DEEP2
galaxy selection. Meeting these goals required us to develop new
empirical techniques to assign colors to our mock galaxies based
purely on the color-environment relation in the DEEP2 data, since
physically motivated galaxy formation models tend not to reproduce
this relation accurately at high redshift.  With the release of this
paper, we also release the mock catalogs for use by the general
public.  Instructions for downloading the mock catalogs and detailed
information on their contents can be found in the Appendix, along with
a specific example showing how these mocks might be useful for
estimating DEEP2 selection effects as a function of halo mass.

Our mock-making technique assigns mock galaxies to the halos and
subhalos in the dark-matter-only simulations by directly imposing the
empirical properties of DEEP2 on the simulations.  We start
by stacking simulation snapshots from different timesteps to construct
lightcones with the geometry of DEEP2 fields, which accurately
reproduces the redshift evolution of cosmic structure.  We then use the
subhalo abundance matching technique to assign galaxy
luminosities to the halos and subhalos in each lightcone.  This
technique posits a tight, monotonic relation between the mass of a
halo and the luminosity of the galaxy it hosts (with the caveat that
the masses of subhalos should be measured before they are accreted
into their parent halos). 

To assign galaxy colors in such a way as to reproduce their strong
dependence on luminosity, redshift, and local environment, we bin the
DEEP2 and mock galaxies in luminosity and redshift and further
subdivide these bins into quintiles of local $n$th-nearest-neighbor
overdensity.  We then draw DEEP2 galaxies at random from these bins
and assign them to the mock galaxies in the corresponding bin, taking
care to account for incompleteness in the DEEP2 sample at faint
magnitudes.  Having produced mock galaxies with redshifts,
luminosities, and colors, we can invert the $k$-correction technique
of W06 to assign apparent $R$-band magnitudes to the mock galaxies.
These allow us to apply the DEEP2 apparent magnitude limit to our mock
sample.  We can also apply a redshift-dependent selection that closely
approximates the effect of the DEEP2 selection in color-color space.
Given these mock DEEP2 targets, we can apply the DEEP2 spectroscopic
target-selection algorithm to reproduce the spatial selection effects
inherent to multiplexed slitmask spectroscopy.  Finally, to account
for observations that fail to yield a redshift, we exclude a fraction
of the mock galaxies according to the color and magnitude-dependent
redshift-failure rate of DEEP2 galaxies.

This mockmaking technique accounts for all major observational effects
that pertain to DEEP2 galaxies.  The resulting mocks accurately
reproduce a wide array of properties of the DEEP2 catalog, including
the luminosity function; the color, magnitude, and redshift
distributions; the color-environment relation and its evolution; and
the inferred DEEP2 HOD for massive halos, provided that the background
cosmology of the simulation is in agreement with current constraints.
This accuracy gives us confidence that our DEEP2 mock catalogs will
allow us to infer the selection of DEEP2 galaxies with reasonable
accuracy as a function of dark-matter-halo mass, as well as the
relations between halo mass and galaxy properties.

In the process of constructing the mock catalogs, we obtained two
results that are of general scientific interest beyond the area of
mock-catalog creation.  First, when assigning galaxy luminosities
using abundance matching techniques, we tried various different assumptions
about the scatter in the mass-luminosity relation and compared the
resulting projected two-point correlation functions $w_p(r_p)$ to the
DEEP2 measurement.  We find that no model with fixed log-normal
scatter (in luminosity at fixed mass) gives simulated $w_p(r_p)$
consistent with the clustering measured in DEEP2.  This is broadly
consistent with the results of \citet{Wetzel10}, who require a 
large value of the scatter to achieve even approximate agreement with
the DEEP2 clustering results.  For reasons discussed in
Section~\ref{sec:scatter}, we do not believe that these issues will
create problems when using these mocks to optimize group-finding
algorithms, although they should be used with care for purposes that
involve understanding the selection of low-mass halos.

In addition, because we have constructed mock catalogs from three
separate N-body simulations with different cosmological parameters, we
can probe the cosmology dependence of our mock-making techniques.  We
find that the HODs in the different mock catalogs vary
strongly with cosmology.  This arises directly from the  technique
for luminosity assignment.  Because abundance matching maps the observed galaxy
luminosity function to the simulated halo mass function at fixed
number density, if the mass function $dn/dM$ in a given simuation has
a lower normalization than the true mass function in the universe,
then the HOD resulting from the galaxy assignment must have a higher
normalization and lower mass cutoff to compensate.  
For cosmologies
with nearly identical mass functions (e.g., points lying along the
same degenerate curve in $\Omega_M$--$\sigma_8$ space), however, the
impact on the HOD will be relatively small.

Thus, it is important when constructing mock catalogs from
$N$-body models to use a simulation whose assumed cosmology is in
agreement with all existing constraints, to ensure the best possible
accuracy in the resulting HOD.  For most applications, we therefore
recommend the mock catalogs we have constructed from the Bolshoi
simulation, since it is consistent with the best cosmological
constraints available.  The mocks constructed from  the L120 and L160
simulations are likely to be generally useful only for testing the
impact of varying assumptions about cosmology on any conclusions drawn
from the mocks.  

Given that the Bolshoi mocks should accurately reproduce the DEEP2
halo selection function, an immediately interesting use to which we
can put them is testing and calibrating an algorithm for detecting
groups and clusters of galaxies.  We have used them for this purpose
in \citet{Gerke12}.

\section*{Acknowledgments}
BFG and RHW were supported by the U.S. Department of Energy under
contract number DE-AC03-76SF00515.  MCC acknowledges the support of
the Spitzer space telescope fellowship program.  We thank Marc Davis,
Jeff Newman, Carlos Frenk, and especially Michael Busha for fruitful
conversations.  We thank Anatoly Klypin and Joel Primack for providing
access to the Bolshoi simulation, which was run on the Pleades machine
at NASA Ames.  We thank Jeremy Tinker for providing us with his code
to compute $w_p(r_p)$ in a simulation box.

\bibliography{ms.bib}

\begin{thebibliography}{}

\bibitem[{Behroozi}, {Conroy}, \& {Wechsler}(2010){Behroozi}, {Conroy}, and
  {Wechsler}]{Behroozi10}
{Behroozi}, P.~S., {Conroy}, C., \& {Wechsler}, R.~H. 2010, \apj, 717, 379

\bibitem[{Behroozi} {et~al.}(2013){Behroozi}, {Wechsler}, {Wu}, {Busha},
  {Klypin}, and {Primack}]{Behroozi11b}
{Behroozi}, P.~S., {Wechsler}, R.~H., {Wu}, H.-Y., {Busha}, M.~T., {Klypin},
  A.~A., \& {Primack}, J.~R. 2013, \apj, 763, 18

\bibitem[{Blaizot} {et~al.}(2005){Blaizot}, {Wadadekar}, {Guiderdoni},
  {Colombi}, {Bertin}, {Bouchet}, {Devriendt}, and {Hatton}]{Blaizot05}
{Blaizot}, J., {Wadadekar}, Y., {Guiderdoni}, B., {Colombi}, S.~T., {Bertin},
  E., {Bouchet}, F.~R., {Devriendt}, J.~E.~G., \& {Hatton}, S. 2005, \mnras,
  360, 159--175

\bibitem[{Bryan} \& {Norman}(1998){Bryan} and {Norman}]{BryanNorman98}
{Bryan}, G.~L., \& {Norman}, M.~L. 1998, \apj, 495, 80

\bibitem[{Busha} {et~al.}(2011){Busha}, {Wechsler}, {Behroozi}, {Gerke},
  {Klypin}, and {Primack}]{Busha10b}
{Busha}, M.~T., {Wechsler}, R.~H., {Behroozi}, P.~S., {Gerke}, B.~F., {Klypin},
  A.~A., \& {Primack}, J.~R. 2011, \apj, 743, 117

\bibitem[Coil {et~al.}(006a)Coil {\em et~al.}]{Coil06a}
Coil, A.~L., et~al. 2006a, ApJ, 638, 668

\bibitem[{Coil} {et~al.}(006b){Coil} {\em et~al.}]{Coil06b}
{Coil}, A.~L., et~al. 2006b, ApJ, 644, 671

\bibitem[Coil {et~al.}(2004)Coil, Newman, Kaiser, Davis, Ma, Kocevski, and
  Koo]{Coil04b}
Coil, A.~L., Newman, J.~A., Kaiser, N., Davis, M., Ma, C.-P., Kocevski, D.~D.,
  \& Koo, D.~C. 2004, ApJ, 617, 765

\bibitem[{Conroy}, {Wechsler}, \& {Kravtsov}(2006){Conroy}, {Wechsler}, and
  {Kravtsov}]{Conroy06a}
{Conroy}, C., {Wechsler}, R.~H., \& {Kravtsov}, A.~V. 2006, \apj, 647, 201--214

\bibitem[Cooper {et~al.}(2006)Cooper {\em et~al.}]{Cooper06a}
Cooper, M.~C., et~al. 2006, MNRAS, 370, 198

\bibitem[Cooper {et~al.}(2007)Cooper {\em et~al.}]{Cooper07}
Cooper, M.~C., et~al. 2007, MNRAS, 376, 1445

\bibitem[Cooper {et~al.}(2005)Cooper, Newman, Madgwick, Gerke, Yan, and
  Davis]{Cooper05}
Cooper, M.~C., Newman, J.~A., Madgwick, D.~S., Gerke, B.~F., Yan, R., \& Davis,
  M. 2005, ApJ, 634, 833

\bibitem[Croton {et~al.}(2006)Croton {\em et~al.}]{Croton06}
Croton, D.~J., et~al. 2006, MNRAS, 365, 11

\bibitem[{Cunha} {et~al.}(2012){Cunha}, {Huterer}, {Busha}, and
  {Wechsler}]{Cunha12}
{Cunha}, C.~E., {Huterer}, D., {Busha}, M.~T., \& {Wechsler}, R.~H. 2012,
  \mnras, 423, 909--924

\bibitem[Davis {et~al.}(2003)Davis {\em et~al.}]{Davis03}
Davis, M., et~al. 2003, Proc. SPIE, 4834, 161

\bibitem[Davis {et~al.}(2007)Davis {\em et~al.}]{AEGIS}
Davis, M., et~al. 2007, ApJ, 660, L1

\bibitem[Davis, Gerke, \& Newman(2004)Davis, Gerke, and Newman]{DGN04}
Davis, M., Gerke, B.~F., \& Newman, J.~A. 2004, astro-ph/0408344

\bibitem[De~Lucia \& Blaizot(2007)De~Lucia and Blaizot]{DB07}
De~Lucia, G., \& Blaizot, J. 2007, MNRAS, 375, 2

\bibitem[Eke {et~al.}(2004)Eke {\em et~al.}]{Eke04}
Eke, V.~R., et~al. 2004, MNRAS, 348, 866

\bibitem[{Faber} {et~al.}(2003){Faber} {\em et~al.}]{Faber03}
{Faber}, S.~M., et~al. 2003, In Instrument Design and Performance for
  Optical/Infrared Ground-based Telescopes. Edited by Iye, Masanori; Moorwood,
  Alan F. M. Proceedings of the SPIE, Volume 4841, pp. 1657-1669 (2003).,
  M.~{Iye} and A.~F.~M. {Moorwood}, eds., volume 4841 of {\em Presented at the
  Society of Photo-Optical Instrumentation Engineers (SPIE) Conference\/}, pp.
  1657--1669

\bibitem[{Faber} {et~al.}(2007){Faber}, {Willmer}, {Wolf}, {Koo}, {Weiner},
  {Newman}, {Im}, {Coil}, {Conroy}, {Cooper}, {Davis}, {Finkbeiner}, {Gerke},
  {Gebhardt}, {Groth}, {Guhathakurta}, {Harker}, {Kaiser}, {Kassin},
  {Kleinheinrich}, {Konidaris}, {Kron}, {Lin}, {Luppino}, {Madgwick},
  {Meisenheimer}, {Noeske}, {Phillips}, {Sarajedini}, {Schiavon}, {Simard},
  {Szalay}, {Vogt}, and {Yan}]{Faber07}
{Faber}, S.~M., {Willmer}, C.~N.~A., {Wolf}, C., {Koo}, D.~C., {Weiner}, B.~J.,
  {Newman}, J.~A., {Im}, M., {Coil}, A.~L., {Conroy}, C., {Cooper}, M.~C.,
  {Davis}, M., {Finkbeiner}, D.~P., {Gerke}, B.~F., {Gebhardt}, K., {Groth},
  E.~J., {Guhathakurta}, P., {Harker}, J., {Kaiser}, N., {Kassin}, S.,
  {Kleinheinrich}, M., {Konidaris}, N.~P., {Kron}, R.~G., {Lin}, L., {Luppino},
  G., {Madgwick}, D.~S., {Meisenheimer}, K., {Noeske}, K.~G., {Phillips},
  A.~C., {Sarajedini}, V.~L., {Schiavon}, R.~P., {Simard}, L., {Szalay}, A.~S.,
  {Vogt}, N.~P., \& {Yan}, R. 2007, \apj, 665, 265--294

\bibitem[{Gerdes} {et~al.}(2010){Gerdes}, {Sypniewski}, {McKay}, {Hao}, {Weis},
  {Wechsler}, and {Busha}]{Gerdes10}
{Gerdes}, D.~W., {Sypniewski}, A.~J., {McKay}, T.~A., {Hao}, J., {Weis}, M.~R.,
  {Wechsler}, R.~H., \& {Busha}, M.~T. 2010, \apj, 715, 823--832

\bibitem[Gerke {et~al.}(2005)Gerke {\em et~al.}]{Gerke05}
Gerke, B.~F., et~al. 2005, ApJ, 625, 6

\bibitem[{Gerke} {et~al.}(2007){Gerke} {\em et~al.}]{Gerke07a}
{Gerke}, B.~F., et~al. 2007, MNRAS, 376, 1425

\bibitem[Gerke {et~al.}(2012)Gerke {\em et~al.}]{Gerke12}
Gerke, B.~F., et~al. 2012, ApJ, 751, 50

\bibitem[{Henriques} {et~al.}(2012){Henriques}, {White}, {Lemson}, {Thomas},
  {Guo}, {Marleau}, and {Overzier}]{Henriques12}
{Henriques}, B.~M.~B., {White}, S.~D.~M., {Lemson}, G., {Thomas}, P.~A., {Guo},
  Q., {Marleau}, G.-D., \& {Overzier}, R.~A. 2012, \mnras, p. 2442

\bibitem[Hogg {et~al.}(2004)Hogg {\em et~al.}]{Hogg04}
Hogg, D.~W., et~al. 2004, ApJ, 601, L29

\bibitem[{Johnston} {et~al.}(2007){Johnston}, {Sheldon}, {Wechsler}, {Rozo},
  {Koester}, {Frieman}, {McKay}, {Evrard}, {Becker}, and {Annis}]{Johnston07}
{Johnston}, D.~E., {Sheldon}, E.~S., {Wechsler}, R.~H., {Rozo}, E., {Koester},
  B.~P., {Frieman}, J.~A., {McKay}, T.~A., {Evrard}, A.~E., {Becker}, M.~R., \&
  {Annis}, J. 2007, arXiv:0709.1159

\bibitem[Kitzbichler \& {White}(2007)Kitzbichler and {White}]{KW07}
Kitzbichler, M.~G., \& {White}, S.~D.~M. 2007, MNRAS, 376, 2

\bibitem[{Klypin} \& {Holtzman}(1997){Klypin} and {Holtzman}]{Klypin97}
{Klypin}, A., \& {Holtzman}, J. 1997, arXiv:astro-ph/9712217

\bibitem[{Klypin}, {Trujillo-Gomez}, \& {Primack}(2011){Klypin},
  {Trujillo-Gomez}, and {Primack}]{Bolshoi}
{Klypin}, A.~A., {Trujillo-Gomez}, S., \& {Primack}, J. 2011, \apj, 740, 102

\bibitem[{Knebe} {et~al.}(2011){Knebe}, {Knollmann}, {Muldrew}, {Pearce},
  {Aragon-Calvo}, {Ascasibar}, {Behroozi}, {Ceverino}, {Colombi}, {Diemand},
  {Dolag}, {Falck}, {Fasel}, {Gardner}, {Gottl{\"o}ber}, {Hsu}, {Iannuzzi},
  {Klypin}, {Luki{\'c}}, {Maciejewski}, {McBride}, {Neyrinck}, {Planelles},
  {Potter}, {Quilis}, {Rasera}, {Read}, {Ricker}, {Roy}, {Springel}, {Stadel},
  {Stinson}, {Sutter}, {Turchaninov}, {Tweed}, {Yepes}, and {Zemp}]{Knebe11}
{Knebe}, A., {Knollmann}, S.~R., {Muldrew}, S.~I., {Pearce}, F.~R.,
  {Aragon-Calvo}, M.~A., {Ascasibar}, Y., {Behroozi}, P.~S., {Ceverino}, D.,
  {Colombi}, S., {Diemand}, J., {Dolag}, K., {Falck}, B.~L., {Fasel}, P.,
  {Gardner}, J., {Gottl{\"o}ber}, S., {Hsu}, C.-H., {Iannuzzi}, F., {Klypin},
  A., {Luki{\'c}}, Z., {Maciejewski}, M., {McBride}, C., {Neyrinck}, M.~C.,
  {Planelles}, S., {Potter}, D., {Quilis}, V., {Rasera}, Y., {Read}, J.~I.,
  {Ricker}, P.~M., {Roy}, F., {Springel}, V., {Stadel}, J., {Stinson}, G.,
  {Sutter}, P.~M., {Turchaninov}, V., {Tweed}, D., {Yepes}, G., \& {Zemp}, M.
  2011, \mnras, 415, 2293--2318

\bibitem[Koester {et~al.}(2007a)Koester {\em et~al.}]{Koester07a}
Koester, B.~P., et~al. 2007a, ApJ, 660, 221

\bibitem[Koester {et~al.}(2007b)Koester {\em et~al.}]{Koester07b}
Koester, B.~P., et~al. 2007b, ApJ, 660, 239

\bibitem[Kravtsov, Klypin, \& Khokhlov(1997)Kravtsov, Klypin, and
  Khokhlov]{Kravtsov97}
Kravtsov, A.~V., Klypin, A.~A., \& Khokhlov, A.~M. 1997, \apj, 111, 73

\bibitem[{Kravtsov} {et~al.}(2004){Kravtsov}, {Berlind}, {Wechsler}, {Klypin},
  {Gottl{\"o}ber}, {Allgood}, and {Primack}]{Kravtsov04}
{Kravtsov}, A.~V., {Berlind}, A.~A., {Wechsler}, R.~H., {Klypin}, A.~A.,
  {Gottl{\"o}ber}, S., {Allgood}, B., \& {Primack}, J.~R. 2004, \apj, 609,
  35--49

\bibitem[{Le F{\`e}vre} {et~al.}(2005){Le F{\`e}vre} {\em et~al.}]{VVDS}
{Le F{\`e}vre}, O., et~al. 2005, \aap, 439, 845--862

\bibitem[{Lilly} {et~al.}(2009){Lilly} {\em et~al.}]{zCOSMOS}
{Lilly}, S.~J., et~al. 2009, \apjs, 184, 218--229

\bibitem[{Muldrew}, {Pearce}, \& {Power}(2011){Muldrew}, {Pearce}, and
  {Power}]{Muldrew11}
{Muldrew}, S.~I., {Pearce}, F.~R., \& {Power}, C. 2011, \mnras, 410, 2617--2624

\bibitem[{Newman} {et~al.}(2012){Newman} {\em et~al.}]{DEEP2}
{Newman}, J.~A., et~al. 2012, ApJ, submitted, arxiv:1203.3192

\bibitem[Peacock \& Smith(2000)Peacock and Smith]{PS00}
Peacock, J.~A., \& Smith, R.~E. 2000, MNRAS, 318, 1144

\bibitem[{Reddick} {et~al.}(2013){Reddick}, {Wechsler}, {Tinker}, and
  {Behroozi}]{Reddick12}
{Reddick}, R.~M., {Wechsler}, R.~H., {Tinker}, J.~L., \& {Behroozi}, P.~S.
  2013, ApJ, in press; arXiv:1207.2160

\bibitem[{Rozo} {et~al.}(2007){Rozo}, {Wechsler}, {Koester}, {Evrard}, and
  {McKay}]{Rozo07}
{Rozo}, E., {Wechsler}, R.~H., {Koester}, B.~P., {Evrard}, A.~E., \& {McKay},
  T.~A. 2007, ArXiv Astrophysics e-prints

\bibitem[{Schechter}(1976){Schechter}]{Schechter76}
{Schechter}, P. 1976, \apj, 203, 297--306

\bibitem[{Tasitsiomi} {et~al.}(2004){Tasitsiomi}, {Kravtsov}, {Wechsler}, and
  {Primack}]{Tasitsiomi04}
{Tasitsiomi}, A., {Kravtsov}, A.~V., {Wechsler}, R.~H., \& {Primack}, J.~R.
  2004, \apj, 614, 533--546

\bibitem[{Trujillo-Gomez} {et~al.}(2011){Trujillo-Gomez}, {Klypin}, {Primack},
  and {Romanowsky}]{TrujiloGomez11}
{Trujillo-Gomez}, S., {Klypin}, A., {Primack}, J., \& {Romanowsky}, A.~J. 2011,
  \apj, 742, 16

\bibitem[{Vale} \& {Ostriker}(2004){Vale} and {Ostriker}]{ValeOstriker04}
{Vale}, A., \& {Ostriker}, J.~P. 2004, \mnras, 353, 189--200

\bibitem[Wetzel \& White(2010)Wetzel and White]{Wetzel10}
Wetzel, A.~R., \& White, M. 2010, \mnras, 403, 1072

\bibitem[Willmer {et~al.}(2006)Willmer {\em et~al.}]{Willmer06}
Willmer, C. N.~A., et~al. 2006, ApJ, 647, 853

\bibitem[{Woo} {et~al.}(2012){Woo}, {Dekel}, {Faber}, {Noeske}, {Koo}, {Gerke},
  {Cooper}, {Salim}, {Dutton}, {Newman}, {Weiner}, {Bundy}, {Willmer}, {Davis},
  and {Yan}]{Woo12}
{Woo}, J., {Dekel}, A., {Faber}, S.~M., {Noeske}, K., {Koo}, D.~C., {Gerke},
  B.~F., {Cooper}, M.~C., {Salim}, S., {Dutton}, A.~A., {Newman}, J., {Weiner},
  B.~J., {Bundy}, K., {Willmer}, C.~N.~A., {Davis}, M., \& {Yan}, R. 2012,
  MNRAS, submitted, arXiv:1203.1625

\bibitem[Yan, White, \& Coil(2004)Yan, White, and Coil]{YWC04}
Yan, R., White, M., \& Coil, A.~L. 2004, ApJ, 607, 739

\bibitem[{Yang}, {Mo}, \& {van den Bosch}(2003){Yang}, {Mo}, and {van den
  Bosch}]{Yang03}
{Yang}, X., {Mo}, H.~J., \& {van den Bosch}, F.~C. 2003, \mnras, 339,
  1057--1080

\bibitem[York {et~al.}(2000)York {\em et~al.}]{SDSS}
York, D.~G., et~al. 2000, AJ, 120, 1579

\bibitem[{Zheng} {et~al.}(2005){Zheng} {\em et~al.}]{Zheng05}
{Zheng}, Z., et~al. 2005, \apj, 633, 791--809

\bibitem[Zheng, Coil, \& Zehavi(2007)Zheng, Coil, and Zehavi]{ZCZ07}
Zheng, Z., Coil, A.~L., \& Zehavi, I. 2007, \apj, 667, 760

\end{thebibliography}

\appendix

\section{The Public Catalogs and Example Applications}
With the publication of this paper, we also release our DEEP2 mock
catalogs for use by the general public.  Mocks for each of the three
different simulation boxes listed in Table~\ref{tab:cones} can be
obtained as binary data tables in FITS format from the World Wide Web at
{\tt http://www.slac.stanford.edu/~risa/deepmocks}.  In this section we
describe the contents of the electronic mock catalog files and give a
few examples of applications for which they might be useful.

\subsection{The catalogs}

We provide two sets of forty lightcones from the Bolshoi simulation,
each of which has the geometry of a single DEEP2 field, with an
additional buffer region around the edge of the field on the sky.  One
set of lightcones is selected to reproduce the primary (high-redshift)
DEEP2 fields, and the other is selected to replicate the EGS.  The
dark-matter halo populations of corresponding lightcones in the two
sets of mocks are identical.  For the L120 and L160 simulations, we
provide two sets of eight lightcones from each box. 
The electronic mock catalogs are stored as binary data tables
conforming to the FITS standard.  They consist of a set of rows, one
for each galaxy, with information stored in columns with the
headings below.  In all cases, a missing value is indicated by the
value $-999$.  

\begin{itemize}
\item {\tt OBJNO}: A unique identification number for each object (galaxy or
  star).  
\item {\tt RA}: Right Ascension.
\item {\tt DEC}: Declination.
\item {\tt Z}: Redshift (including the peculiar-velocity Doppler shift).
\item {\tt MAGR}: The apparent R-band magnitude $m_R$.
\item {\tt ABSBMAG}: The absolute B-band magnitude $M_B$.
\item {\tt UB\_0}: The rest-frame $U-B$ color.
\item {\tt ZQUALITY}: Analagous to the DEEP2 redshift-quality flag
  $Q$, except that we do not use all of the the DEEP2 confidence codes
  from Q=1 to 4.  Galaxies that are assigned successful redshifts in
  our mock-observation algorithms have $Q=4$, targeted galaxies that
  had redshift failures in our algorithm have $Q=1$, stars have
  $Q=-1$, and galaxies not targeted for spectroscopy have $Q=-2$.
\item {\tt SELECTED}: Galaxies that passed the redshift selection cut
  in Equation~\ref{eqn:zselect} (or the weighted redshift selection we
  apply in the EGS mocks) have this field set to unity.
\item {\tt MASKREGION} Galaxies that fall in the region used for DEEP2
  maskmaking, rather than the buffer region around the edge, have this
  field set to unity.   Limiting to these galaxies and the ones
  flagged in the previous item is important for estimating the
  DEEP2 selection function in many cases.
\item {\tt PGAL}  Analagous to the DEEP2 star-galaxy separation
  probability.  Mock galaxies have this field set to unity; stars have
  it set to zero.
\item {\tt XPOS, YPOS,} and {\tt ZPOS} The $xyz$ position of this galaxy
  in the simulation box (in units of the box size).
\item {\tt VR} The velocity of this galaxy along the line of sight
  (where positive velocity implies movement away from the observer).
  This can be combined with the {\tt Z} value to compute the
  cosmological redshift.
\item {\tt GROUPID} A unique identifier for the distinct (parent) halo
  to which this galaxy belongs.
\item {\tt HALOID} A unique identifier for the subhalo to which this
  galaxy belongs.
\item {\tt MASS}\footnote{This column is named {\tt M180} in the L120
    and L160 mocks, where it gives the mass  computed within a sphere whose
    radius is that at which the subhalo density is 180 times 
  the background density.} The virial mass of this galaxy's dark-matter
  subhalo, as defined in \citet{BryanNorman98}.
\item {\tt VMAX} The maximum circular velocity of the subhalo (at the
  present time).
\item {\tt GROUPMASS} The mass of this galaxy's parent halo
\item {\tt CENTRAL} Galaxies that are the central galaxies of their
  parent halos have this field set to unity.
\item {\tt SIGMA7\_RAW} Seventh-nearest neighbor surface density
  $\Sigma_7$ for
  this galaxy, computed before applying DEEP2 selection
  cuts. 
\item {\tt DELTA7\_RAW} Seventh-nearest neighbor overdensity
  $\delta_7 = \Sigma_7/\langle\Sigma_7(z)\rangle$
\item {\tt SIGMA3\_OBS} Third-nearest neighbor surface density
  $\Sigma_3$ for
  this galaxy, computed after applying DEEP2 selection
  cuts.
\item {\tt DELTA3\_OBS} Third-nearest neighbor overdensity
  $\delta_3=\Sigma_3/\langle\Sigma_3(z)\rangle$. 
\item {\tt ECOMDIST} Comoving distance to the edge of the survey in
  comoving $h^{-1}$ Mpc.  If this value is less than 1$h^{-1}$ Mpc,
  the value of $\delta_3$ for this galaxy is likely to be inaccurate.
\item {\tt EANGDIST} Angular distance to the edge of the survey.
  (This column and the previous one are only defined for observed
  galaxies.)
\item {\tt WEIGHT} The incompleteness weight $w_\mathrm{corr}$ assigned to this galaxy.
\end{itemize}
\subsection{The mass dependence of DEEP2 selection}

\begin{figure}
\centering
\includegraphics[width=3.25in]{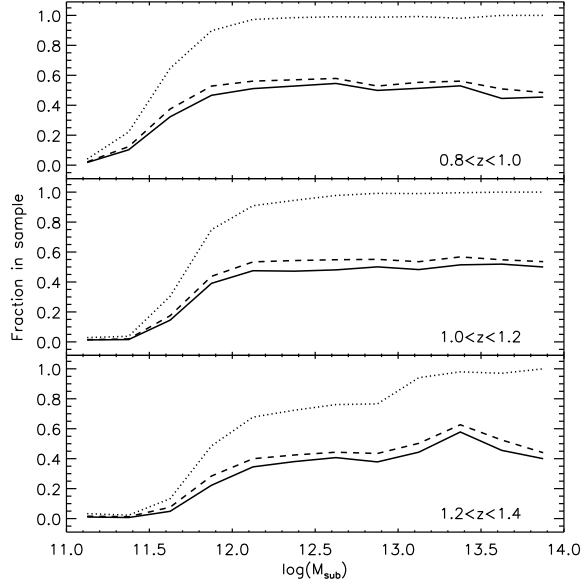}
\caption{The selection function of mock DEEP2 galaxies versus subhalo
  mass, in bins of redshift.  Solid lines show the fraction of
  subhalos that conain a galaxy that was included in the sample of
  observed mock galaxies with successful redshifts in the mocks.
  Dashed lines show the fraction that were scheduled for observation
  by the DEEP2 spectroscopic targeting algorithm.  Dotted lines show
  the fraction that exceed the DEEP2 $R=24.1$ apparent magnitude
  limit. }
\label{fig:selfn}
\end{figure}

One of our goals in producing mock catalogs was to understand the
selection probability of DEEP2 galaxies as a function of their halo
mass.  As we discussed in Sections~\ref{sec:scatter}, the overall
mass-selection function in the mocks is likely to have a low-mass
cutoff that is sharper than is the case in the real universe, owing to
the zero-scatter abundance matching model we used to populate the
N-body model with galaxies; however, it is possible that other
mass-dependent selection effects will dominate, making the precise
shape of the HOD cutoff less relevant.  At any rate, because various
DEEP2 selection effects, including targeting and redshift-failure
effects, depend on local galaxy density, color, and apparent
magnitude, one may well be concerned that the DEEP2 selection function
might be a strong function of halo mass, beyond the cutoff at low mass
from the magnitude limit.  We can investigate this in detail with the
mock catalogs produced in this study.

Figure~\ref{fig:selfn} shows the fraction of mock galaxies that passed
various DEEP2 selection cuts, in three different redshift ranges, as a
function of subhalo mass.  Dotted lines show galaxies with $R\le
24.1$, dashed lines show galaxies that were assigned to spectroscopic
slits by the slitmask-making algorithm, and solid lines show galaxies
that were successfully assigned redshifts.  To produce these curves,
we selected galaxies with {\tt SELECTED}, {\tt MASKREGION}, and {\tt
  PGAL} values equal to unity (to exclude galaxies that had no
probability of selection); then we computed the fraction that had {\tt
  MAGR} $\le 24.1$ and {\tt ZQUALITY} greater than zero and equal to
4, respectively.  The figure shows a broad mass cutoff in the
mocks, spanning nearly an order of magnitude in mass at $z\sim 1$.
The breadth of this transition occurs mostly because the DEEP2 flux
limit translates into a strongly color-dependent luminosity cut, as
shown in Figure~\ref{fig:colmag_compare}. This transition is probably
somewhat broader in reality, since the mock HOD cuts off too sharply
at low masses; however, the dominant effect is the variable absolute
magnitude limit, not the shape of the HOD, since the cutoff in the HOD
(\emph{cf.} Figure~\ref{fig:hod_compare}) is sharper than the
cutoff in mass shown in Figure~\ref{fig:selfn}.

The mass cutoff also broadens further at higher
redshift, such that the magnitude limit alone introduces some level of
incompleteness all the way up to $M\sim 10^{13.5}$, the mass scale of
massive groups, at the highest redshifts in DEEP2.  It is
important to note, though, that this figure considers only the
detection probability of the single galaxy at the center of each halo
or subhalo.  Massive group and cluster halos, containing multiple
galaxies, might still have their satellites detected, even if their
centrals are not, so the selection function for groups will look
somewhat different than this \citep{Gerke12}.

The mass dependence of the DEEP2 targeting and redshift failures is
much weaker than the magnitude selection.  There is a slight drop
in the spectroscopic targeting rate at high masses, particularly at
high redshift.  This can probably be
attributed to the crowding of slits on DEIMOS slitmasks, which will be
somewhat more severe for galaxies in groups and clusters.  However, as
discussed in several other papers \citep{Cooper05, Gerke05, Gerke12},
such crowding effects on the sky do not necessarily translate into
strong selection effects in three-space; hence, the impact on the
mass-selection function here is quite weak.  There is also a 
slight increase in the redshift-success rate with mass, corresponding
to the drop in redshift success for the faintest apparent magnitudes.
However, the main conclusion from this figure is that, for most
applications, the targeting and redshift-failure rates in DEEP2 can
safely be approximated as flat functions of mass.

\end{document}